\documentclass[aps,prc,showpacs,showkeys,twocolumn,nofootinbib]{revtex4}

\usepackage{graphicx,epsfig,amsmath,amssymb}
\usepackage[bookmarksnumbered,bookmarksopen]{hyperref}
\usepackage[capitalize]{cleveref}
\usepackage[utf8]{inputenc}

\begin{document}

\title{``Snowballs in hell'': Microscopic study of deuteron production in PbPb collisions at $\sqrt{s} = 2.76$ TeV via hydrodynamics and hadronic afterburner}

\author{Dmytro~Oliinychenko$^{1}$, Long-Gang~Pang$^{1,2}$, Hannah~Elfner$^{3,4, 5}$ and Volker Koch$^{1}$}

\affiliation{$^1$Lawrence Berkeley National Laboratory, 1 Cyclotron Rd, Berkeley, CA 94720, US}
\affiliation{$^2$Physics Department, University of California, Berkeley, CA 94720, USA}
\affiliation{$^3$Frankfurt Institute for Advanced Studies, Ruth-Moufang-Strasse 1, 60438 Frankfurt am Main, Germany}
\affiliation{$^4$Institute for Theoretical Physics, Goethe University, Max-von-Laue-Strasse 1, 60438 Frankfurt am Main, Germany}
\affiliation{$^5$GSI Helmholtzzentrum für Schwerionenforschung, Planckstr. 1, 64291 Darmstadt, Germany}
\date{\today}

\begin{abstract}

The deuteron yield in Pb+Pb collisions at $\sqrt{s_{NN}} = 2.76$~TeV is consistent with thermal production at a freeze-out temperature of $T = 155$~MeV. The existence of deuterons with binding energy of 2.2~MeV at this temperature was described as ``snowballs in hell'' [P.~Braun-M\"{u}nzinger, B. D\"{o}nigus and N. L\"{o}her,
  CERN Courier, August 2015]. We provide a microscopic explanation of this phenomenon, utilizing relativistic hydrodynamics and switching to a hadronic afterburner at the above mentioned temperature of $T = 155$~MeV. The measured deuteron $p_T$-spectra and coalescence parameter $B_2(p_T)$ are reproduced without free parameters, only by implementing experimentally known cross-sections of deuteron reactions with hadrons, most importantly $\pi d \leftrightarrow \pi n p$.
\end{abstract}

\pacs{25.75.Dw}
\keywords{Heavy ion collisions, deuteron, transport approach, relativistic hydrodynamics}
\maketitle


\section{\label{sec:Intro}Introduction}

Light nuclei production was measured recently by the ALICE collaboration in Pb+Pb collisions at $\sqrt{s_{NN}} = 2.76$ TeV \cite{Adam:2015vda}. At first glance it may seem surprising that light nuclei with binding energies of a few MeV are produced at all in such violent collisions \cite{Cern_courier}. A number of models provide explanations \cite{Kapusta:1980zz}, two of them being particularly popular: thermal production \cite{Siemens:1979dz,Andronic:2010qu,Andronic:2012dm,Cleymans:2011pe,Oliinychenko:2016dtb} and coalescence \cite{Sato:1981ez,Gutbrod:1988gt,Mrowczynski:1992gc,Csernai:1986qf,Polleri:1997bp,Mrowczynski:2016xqm,Bazak:2018hgl,Sun:2016rev,Dong:2018cye,Sun:2018jhg}. The thermal model assumes perfect chemical equilibrium above the chemical freeze-out temperature $T_{CFO}$ and a sharp chemical freeze-out of all hadrons at $T_{CFO}$. Below this temperature the yields are unchanged, but particles can scatter elastically until the system reaches the kinetic freeze-out temperature $T_{KFO}$. The yields measured by ALICE are described very well by $T_{CFO} = 155$ MeV, while the transverse momentum spectra are described by a kinetic freeze-out at the temperature $T_{KFO} = 115$ MeV \cite{Adam:2015vda}. In contrast to this view, the coalescence approach postulates that light nuclei are formed only at late times of the fireball evolution, by binding nucleons that reside close in phase space. Therefore, the coalescence model predicts momentum spectra of nuclei with number of protons $Z$ and number of neutrons $A - Z$ being proportional to the powers of the proton and neutron spectra $\left(E_p \frac{dN_p}{d^3p}\right)^Z \left(E_n \frac{dN_n}{d^3p}\right)^{A-Z}$. Even though the thermal and coalescence models adopt different assumptions, they predict similar deuteron yields \cite{Mrowczynski:2016xqm}. It was proposed to compare $^4$He and $^4$Li production in heavy ion collisions to distinguish the two approaches \cite{Mrowczynski:2016xqm,Bazak:2018hgl}, but the measurement has not been performed yet.

The deuteron yield at midrapidity $\frac{dN_d}{dy}|_{y=0}$, measured by ALICE \cite{Adam:2015vda}, is described very well by the 
thermal model via

\begin{eqnarray} \label{Eq:I}
\frac{dN_d}{dy}|_{y=0} = \frac{g_d V_{CFO}}{2\pi^2 \hbar^3} T_{CFO} \, m_d^2 \,
 K_2\left(\frac{m_d}{T_{CFO}}\right) \,,
\end{eqnarray}

where $g_d = 3$ is deuteron degeneracy, $V_{CFO}$ is the volume of the fireball at the hadronic chemical freeze-out, $m_d = 1.8756$ GeV is the deuteron mass and $T_{CFO} = 155$ MeV is the chemical freeze-out temperature. The thermal model assumes that this yield is unchanged on average during the fireball evolution after the hadronic chemical freeze-out. This assumption is non-trivial, because collisions in the hadronic phase can still easily destroy the deuteron. A justification at low energies is provided by~\cite{Siemens:1979dz}, which assumes that the reaction $d N \leftrightarrow N np $ reaches equilibration, so deuterons are both destroyed and produced at equal rates. This assumption allows to relate the deuteron to proton yields ratio $R_{dp}$ to the entropy of nucleons per nucleon $S_N$ via a simple relation:
\begin{eqnarray} \label{Eq:Siemens_eq3}
  S_N = 3.95 - \ln R_{dp} \,.
\end{eqnarray}
In Ar+KCl and Ne+NaF collisions at $E_{lab} = 400$, 800, and 2100 MeV~\cite{Nagamiya:1981sd}, analyzed in~\cite{Siemens:1979dz}, both the baryon number and the entropy are carried mostly by nucleons. Therefore, the entropy of nucleons per nucleon $S_N$ is approximately equal to {\em total} entropy per baryon. Assuming ideal hydrodynamic expansion, where the entropy per baryon is conserved, one concludes that the $d/p$ ratio is constant during the expansion too. These considerations were criticized already at low energies for neglecting the impact of deuterons to the total entropy and total baryon number~\cite{Biro:1981es}, as well as for neglecting contributions of mesons and resonances to the entropy, and also for neglecting the effect of quantum statistics~\cite{Stoecker:1981kf,Csernai:1987ri}.

At LHC energies, where the ALICE measurement was taken and  which are 3 orders of magnitude higher, this justification has to be completely revisited. The system at this energy is meson-dominated, while at low energies it is baryon-dominated. The main reaction producing deuterons is different: as we show below, it is $ \pi np \leftrightarrow \pi d$. This reaction leads to the same relation between the deuteron to proton ratio and the entropy per nucleon carried by nucleons as in Eq. (\ref{Eq:Siemens_eq3}). However, at LHC energies neither the entropy, nor the baryon number are carried mostly by nucleons. Therefore there is no reason to assume that $S_N$ is conserved. Collisions of nucleons with pions, resonances, and $B\bar{B}$ annihilations distort the explanation of \cite{Siemens:1979dz} applied at LHC energies. Here we show that the assumption of approximately constant deuteron yield during the afterburner evolution is actually justified at ALICE energies, but the model from \cite{Siemens:1979dz} is hardly applicable there.

To gain more insight into this question, we study deuteron production in a hybrid approach, i.e. relativistic hydrodynamics + non-equilibrium hadronic transport. This approach has been well established for the dynamical description of heavy ion collisions at high energies \cite{Hirano:2012kj,Petersen:2008dd,Werner:2010aa,Ryu:2012at}, but even within hybrid approaches light nuclei spectra are typically computed via coalescence of the final protons and neutrons (see for example \cite{Sombun:2018yqh}). In pure transport calculations for lower energies coalescence is also applied (e.g. \cite{Gyulassy:1982pe}). A notable exception is an old work \cite{Danielewicz:1991dh}, where the deuteron is treated in a pure transport model as a pointlike particle and is allowed to scatter. However, this work is applied to low energies, where different deuteron-producing reactions are dominant. In a Ref. \cite{Oh:2009gx} a transport approach with a similar deuteron treatment is adopted to simulate Au+Au collisions at $\sqrt{s_{NN}} = 200$ GeV. However, in this work the $NN \leftrightarrow \pi d$ production channel is assumed to be dominant, which contradicts experimental data, see Fig. \ref{Fig:I}. 

Here we simulate deuterons as pointlike particles in a hybrid approach, producing them thermally from hydrodynamics at particlization and allowing them to subsequently rescatter in the hadronic phase. As in works \cite{Danielewicz:1991dh,Oh:2009gx,Longacre:2013apa} we treat the deuteron in the hadronic phase as a pointlike particle, which allows to study its time evolution. This allows to evaluate the assumptions of the thermal and coalescence models from the underlying non-equilibrium transport.
Indeed, the transport approach is more fundamental than both of the models --- both can be derived from it under certain assumptions. Thermal yields follow from transport, if one assumes full chemical equilibration and a rapid chemical freeze-out of all hadrons and light nuclei at a constant-temperature hypersurface. The coalescence model follows from transport, if one assumes that deuterons are mainly produced via $Xnp \leftrightarrow Xd$ reactions (here $X$ is an arbitrary hadron), which remain equilibrated until late times. Then the forward and reverse rates should be equal and  therefore the phase-space distribution equals $\mathit{f}_d(\vec{p}) = \mathit{f}_n(\vec{p}_n) \mathit{f}_p(\vec{p}_p) \delta(\vec{p}-\vec{p_p}-\vec{p_n})$. Accounting for the wavepacket nature of the particles one then obtains Eq. (1) from the coalescence approach \cite{Sun:2018jhg}. We show that the underlying assumptions of the coalescence model are rather well-justified: at low collision energies $Nd \leftrightarrow N np$ reactions are dominating, at high energies --- $\pi d \leftrightarrow \pi np$ reactions. Both have cross-sections, which are higher than typical inelastic hadronic cross-sections; this property allows to equilibrate these reactions fast and keep them equilibrated for a sufficiently long time until kinetic freeze-out. On the contrary, we find that the thermal model assumptions are not fulfilled: deuteron chemical freeze-out occurs later than for most hadrons. Moreover, the chemical and kinetic freeze-out of deuteron approximately coincide in time, because, unlike for most of the hadrons, the d+hadrons inelastic cross-sections are typically larger than the elastic ones. Despite the assumptions of the thermal model being violated, it fits the deuteron yield very well. This is surprising and serves as a part of the motivation of this study. We suggest a possible solution in Sections \ref{Sec:deuteron_yield} and \ref{Sec:toy_model}.

In this paper we concentrate exclusively on deuteron production, because our approach becomes increasingly harder for heavier nuclei. The cross-sections of deuteron interactions are adjusted to match the experimental data. These cross-sections, along with the general description of the simulation are given in Section \ref{Sec:Methodology}, while technical parts of the implementation can be found in Appendices \ref{AppendixA} and \ref{AppendixB}. Deuteron multiplicities, reaction rates and freeze-out are covered in Section \ref{Sec:Results}, followed by the discussion and outlook in Section \ref{Sec:Discussion}.

\section{Methodology} \label{Sec:Methodology}

In this work we constrain ourselves to simulating central, 0-10\%, Pb+Pb collisions at $\sqrt{s_{NN}} = 2.76$ TeV, which were measured by the ALICE collaboration. We utilize a well-established hybrid (hydrodynamics + transport) approach, which includes: generating initial condition for hydrodynamics, solving (3+1)D partial differential relativistic viscous hydrodynamical equations, particlization - sampling hadrons on the constant temperature hypersurface, and simulating subsequent hadronic rescatterings via a transport approach. The latter is also called hadronic afterburner. The main subject of our interest is the microscopic evolution of deuterons during this last  afterburner stage.


\subsection{Hydrodynamics}

The (3+1)D relativistic hydrodynamic equations are solved using CLVisc \cite{Pang:2018zzo} to simulate the fluid dynamic evolution of strongly coupled quark gluon plasma produced in $\sqrt{s_{NN}} = 2.76$ TeV Pb+Pb collisions at the LHC. 
The initial entropy density distributions in Pb+Pb collisions are given by the Trento Monte Carlo model \cite{Bernhard:2016tnd} with default parameters to approximate the IP-Glasma initial condition \cite{Schenke:2012wb}.
The initial conditions and transport coefficients such as the ratio between shear viscosity over entropy density $\eta/s$ have been tuned in event-by-event CLVisc simulations to fit the experimental data in $0-5\%$ most central collisions.
It has been verified that centrality classes determined by the total entropy in the initial state of hydrodynamic simulations describe experimental data well at other centralities.
In the present study, we first simulate 2000 collisions in $0-10\%$ centrality to get the event-by-event initial entropy density distributions using Trento. We then align all the events by shifting to their centers of mass and rotating to their second order participant planes before event averaging to produce a smooth one-shot initial condition. 

The computationally less intensive single-shot smooth evolution is sufficient for our purposes, since the thermal density ratios between different particle species are mainly determined by the mass, the spin-degeneracies of the hadrons and the freeze-out temperature in the comoving frame of the fluid. The total yields of hadrons and deuterons are hardly affected by the fluid velocity, while the transverse momentum and the anisotropic azimuthal angle distributions are sensitive to the fluid velocity on the hypersurface. The one-shot relativistic hydrodynamic simulation is sufficient to provide the fluid velocity profile on the hypersurface. Instead of performing event-by-event hydrodynamic calculations, we have simulated the hadronic afterburner 10000 times with different samples of hadrons and deuterons from the hyper-surface, to provide good statistics for the deuteron and hadron yields.

The hydrodynamic evolution starts at $\tau_0=0.3$ fm with a shear viscosity over entropy density ratio $\eta_v/s=0.16$. The expansion rate of the quark gluon plasma is driven by the pressure gradient from the s95p-pce lattice equation-of-state \cite{Borsanyi:2012ve}, which matches a chemically equilibrated hadron gas at temperatures between 150 and 184 MeV. On the constant temperature hyper-surface with $T_{\rm frz}=155$ MeV, which is equal to the chemical freeze-out temperature $T_{CFO}$ in the thermal model, we sample hadrons as well as deuterons using Cooper-Frye formula \cite{Cooper:1974mv},
\begin{equation}
    \frac{dN_i}{dYp_T dp_T d\phi} = \frac{g_i}{(2\pi)^3} \int p^{\mu}d\Sigma_{\mu} f_{\rm eq}(1 + \delta f)
    \label{eq:cooper_frye}
\end{equation}
where $d\Sigma_{\mu}$ is the freeze-out hyper-surface element determined by $T_{\rm frz}$,
$g_i$ is the spin degeneracy of particle $i$.
Particles passing through the freeze-out hyper-surface elements are assumed to obey 
Fermi-Dirac (for baryons) and Bose-Einstein distributions (for mesons) with the non-equilibrium correction $\delta f$
to the equilibrium distribution $f_{\rm eq}$,

\begin{eqnarray}
    f_{\rm eq} &=& \frac{1}{\exp\left[(p\cdot u )/T_{\mathrm{frz}} \right] \pm 1 } \\
    \delta f &=& (1 \mp f_{\rm eq})\frac{p_{\mu} p_{\nu} \pi^{\mu\nu}}{2 T_{\mathrm{frz}}^2 (\varepsilon + P)}
    \label{eqn:feq}
\end{eqnarray}
where $p$ are the four-momenta of the sampled hadrons, $\pm$ is for fermion/bosons, respectively. $\varepsilon$, $P$, $u$ and $\pi^{\mu\nu}$ are the local energy density, pressure, fluid four-velocity and shear stress tensor given by dissipative hydrodynamic simulations. The non-equilibrium corrections improve the spectra at high transverse momenta, but have very small effect on the yields.
The deuterons are treated as normal hadrons with mass 1.8756 GeV, spin 1 and baryon number 2.

\subsection{Hadronic afterburner} \label{sec:hadronic_afterburner}

\begin{figure}[tbh]
\includegraphics[width=0.5\textwidth]{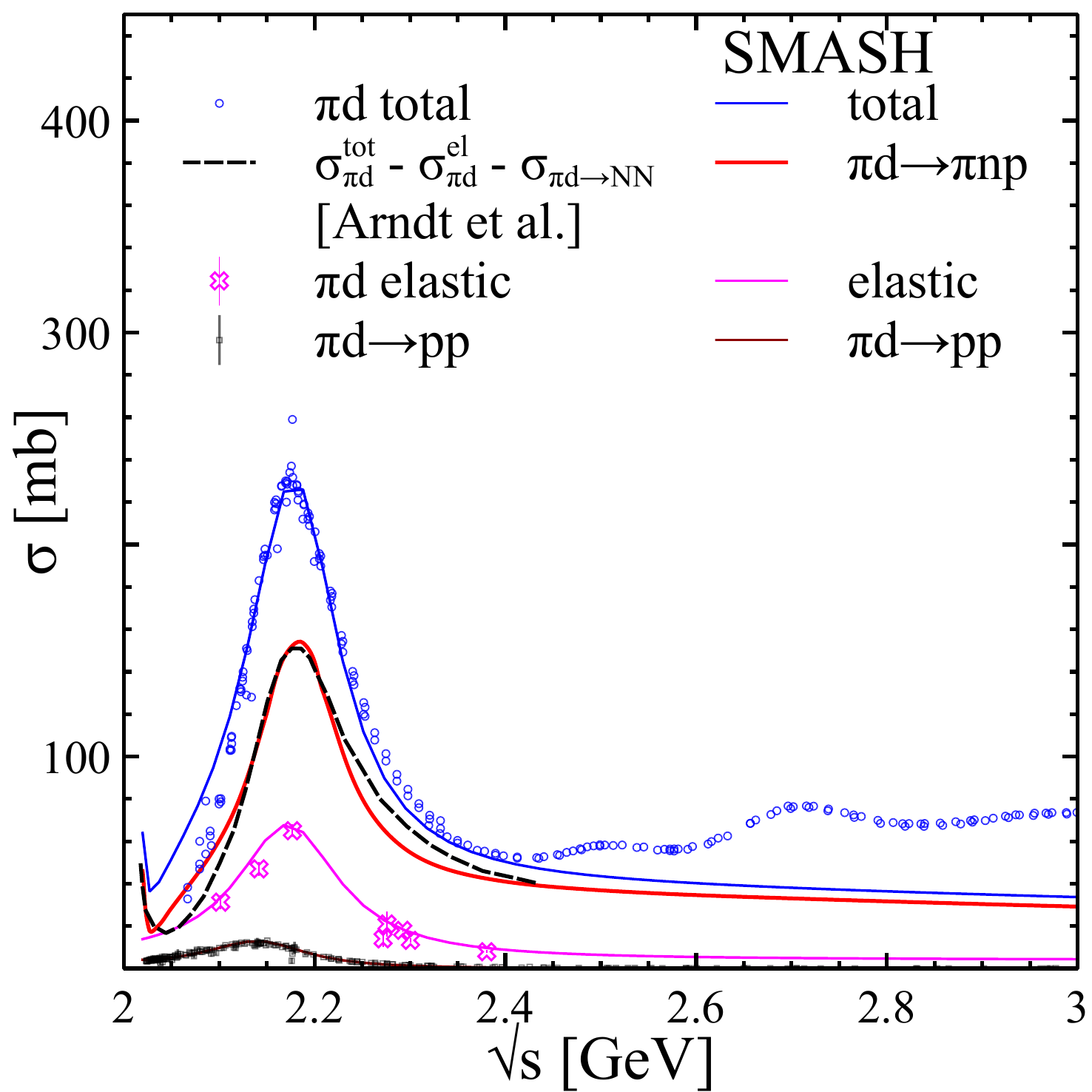}
\caption{Deuteron-pion interaction cross-sections from SAID database \cite{SAID_database} and partial wave analysis \cite{Arndt:1994bs} are compared to our parametrizations (Tables \ref{Tab:I} and \ref{Tab:II} in the appendix). Inelastic $d\pi \leftrightarrow np \pi$ reactions are the most important for deuteron production and disintegration at high energies. The large total $\pi d$ cross-section is responsible for the late deuteron freeze-out. Also, the inelastic $\pi d$ cross-section is larger than elastic. As a consequence, for deuteron  chemical and kinetic freeze-out coincide - see Fig. \ref{Fig:VI}.}
\label{Fig:I}
\end{figure}

\begin{figure*}
\includegraphics[width=0.49\textwidth]{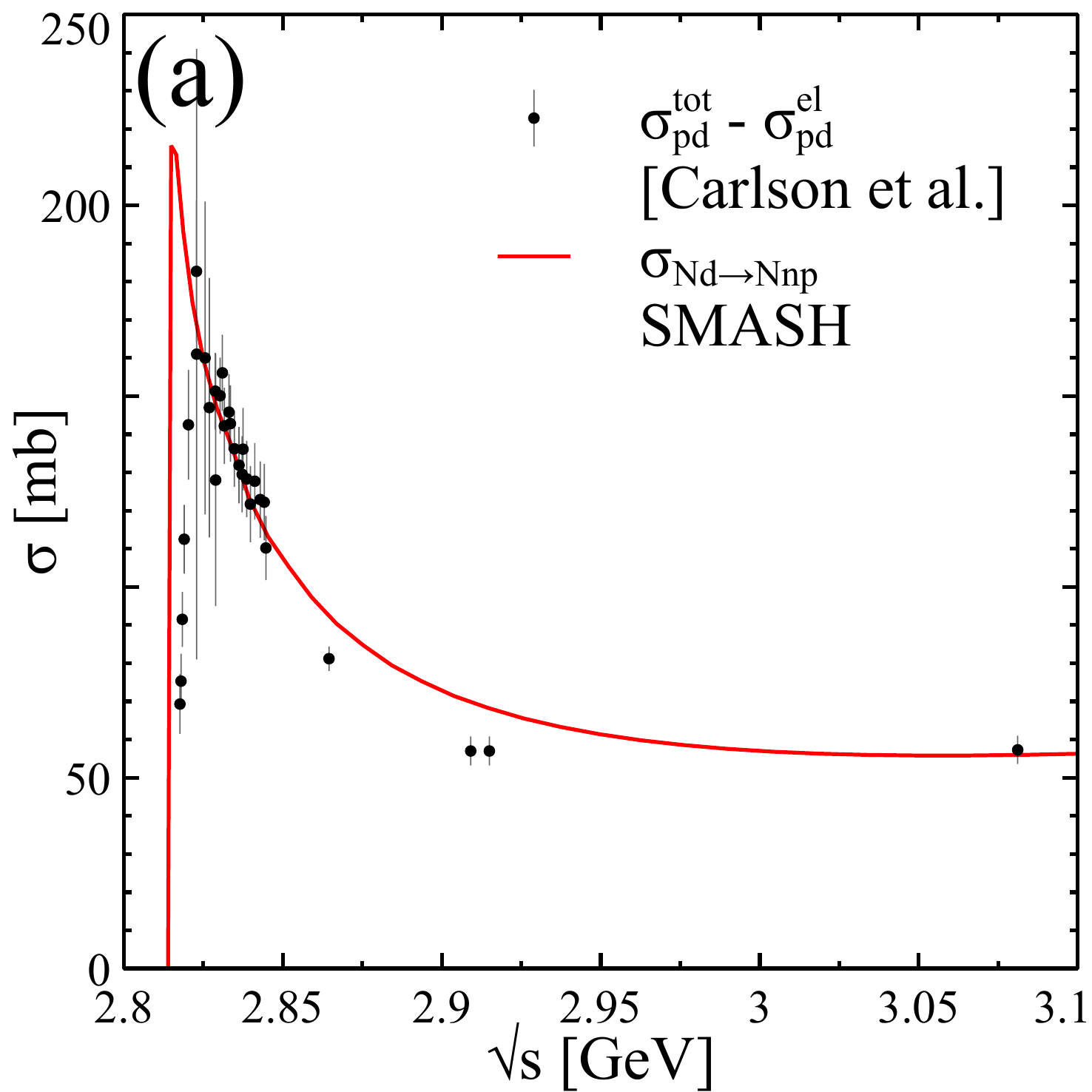}
\includegraphics[width=0.49\textwidth]{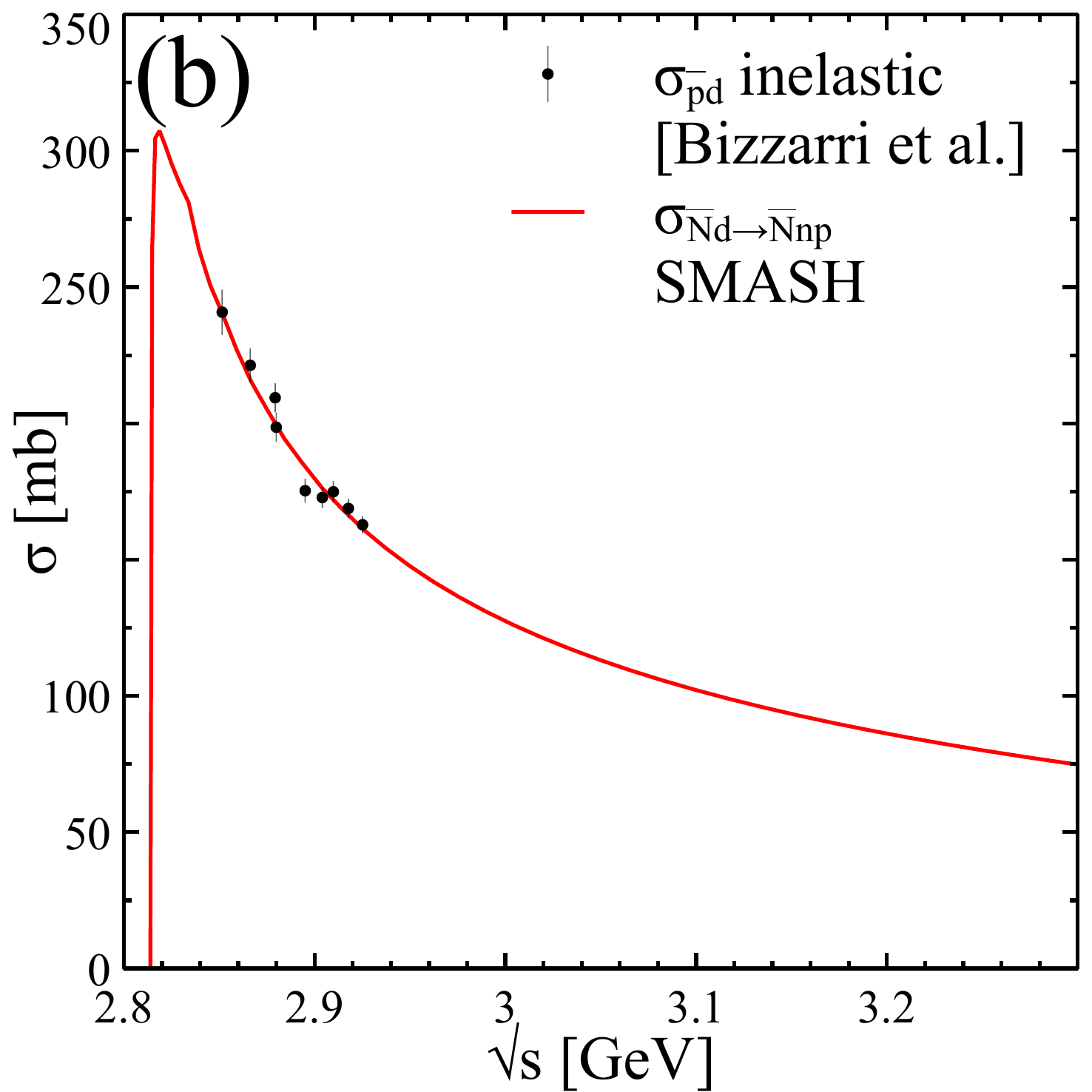}
\caption{Deuteron disintegration cross-sections by nucleons (a) and anti-nucleons (b) as given by Tabs. \ref{Tab:I} and \ref{Tab:II}, compared to data \cite{Bizzarri:1973sp,Carlson:1973}. Although the cross-sections are large, these processes are of minor importance compared to $\pi d \leftrightarrow \pi np$, because pions are much more abundant in Pb+Pb collisions at $\sqrt{s} = 2.76$ TeV.}
\label{Fig:II}
\end{figure*}

The recently developed SMASH transport approach \cite{Weil:2016zrk} serves us as the afterburner. We perform simulations with the full SMASH table of hadrons, where most of the hadron resonances listed in the Particle Data Group collection \cite{Patrignani:2016xqp} are included. Hadronic interactions within SMASH encompass: elastic collisions; resonance formations and decays; $2 \to 2$ inelastic reactions such as $NN \to N\Delta$, $NN \to N N^*$, $NN \to N\Delta^*$ ($N^*$ and $\Delta^*$ denote all nucleon- and delta-resonances), and strangeness exchange reactions; soft string formation and decay into multiple hadrons. The main update relevant for this study since the publication \cite{Weil:2016zrk} is the high-energy hadronic interactions via string formation, including baryon-antibaryon annihilations. All the reactions, except the ones with strings (which include baryon-antibaryon annihilation), obey the detailed balance principle. The implementation of hadronic interactions in SMASH is described in detail in \cite{Weil:2016zrk}, while \cite{Steinberg:2018jvv} is devoted specifically to reactions involving strangeness. Soft string formation and fragmentation, including baryon-antibaryon annihilations, are adapted from the UrQMD code \cite{Bass:1998ca} and will be described in detail elsewhere. The main difference to the UrQMD implementation is that the Lund fragmentation functions from Pythia are employed for newly produced particles. For this study we use the code version SMASH 1.4. We treat the deuteron as a pointlike particle, as any hadron in SMASH. This approach was also adopted in \cite{Oh:2009gx,Danielewicz:1991dh,Longacre:2013apa}. Treating deuterons as pointlike particles is only justified, when the mean free path is at least twice larger than the deuteron size. We find that in our simulation this condition is fulfilled only starting from time $t \gtrsim 25$    fm/c. At earlier time our deuterons are not defined as particles and should be understood as correlated nucleon pairs. We implement all the reactions with deuterons such that they obey detailed balance principle (for details see the Appendix).

\begin{table}
\begin{tabular}{ccc}
\toprule
$X$          &     $\sigma_{d+X}^{\mathrm{inel}}$ [mb]   &  Refs. \\
$\pi^{\pm}$  &     80 - 160   &  \cite{Arndt:1994bs}  \\
$K^+$        &     $<$ 40      &  \cite{Carlson:1973}  \\
$K^-$        &     $<$ 80      &  \cite{Sibirtsev:2006yw,Giacomelli:1972vb}  \\
$p$          &    50 - 100    &  \cite{Bugg:1968zz}  \\
$\bar{p}$    &    80 - 200    &  \cite{Bizzarri:1973sp}  \\
\botrule
\end{tabular}
\caption{Inelastic d + X cross-sections in the range of $\sqrt{s} - \sqrt{s_{thr}} = [0.05, 0.25]$ GeV. This is the most relevant range for our afterburner simulation. Pion + deuteron inelastic reactions are the most important at ALICE energies, but not because the cross-section is large, rather because pions are so abundant.}
\label{Tab:0}
\end{table}

Next let us sort the $d+X$ reactions, where $X$ denotes any hadron, by importance for deuteron production at $\sqrt{s_{NN}} = 2.76$ TeV. The larger the product of the $d+X$ cross-section and the density of hadron $X$, the more frequent the $d+X$ reaction occurs and hence the more important it is. At $\sqrt{s_{NN}} = 2.76$ TeV the most abundant hadrons produced in the PbPb collision at midrapidity are pions: for comparison, the midrapidity yields of pions, kaons and (anti)protons in the central collisions, measured by ALICE are $\frac{dN}{dy}^{\pi^{\pm}} = 733 \pm 54$, $\frac{dN}{dy}^{K^{\pm}} = 109 \pm 9$ and $\frac{dN}{dy}^{p} = 33 \pm 3$ \cite{Abelev:2012wca}. Typical hadron-hadron collisions in the afterburner occur in the range of hadron-hadron center of mass energies $\sqrt{s} -\sqrt{s_{thr}} = (0.5 - 2) T$, where $T$ is the temperature. For these $\sqrt{s}$ the cross-sections are summarized in Tab. \ref{Tab:0}. One can see that the inelastic cross-sections of the deuteron with (anti)nucleons are as high as with pions, however, pions are about 20 times more abundant. Therefore, the number of inelastic reactions $d+\pi$ in the afterburner is expected to be at least an order of magnitude larger than $d+K$, $d+N$ and $d+\bar{N}$ combined. In contrast, at low energies of heavy ion collisions, where the proton yield is larger than the pion yield, $Nd \leftrightarrow Nnp$ reactions dominate the deuteron disintegration and production. We have implemented $\pi d$ reactions with cross-sections as shown in Fig. \ref{Fig:I}, where the $\pi d \leftrightarrow \pi pn$ is assumed to be responsible for the inelastic cross-section. This assumption is experimentally well-justified in the relevant $\sqrt{s}$ region \cite{Arndt:1994bs}. For the purpose of this study implementing deuteron interaction with pions would be enough. However, to test the above estimate and for the future application at lower energies, we have also implemented $N d \leftrightarrow N n p$ and $\bar{N} d \leftrightarrow \bar{N} n p$ (see Fig. \ref{Fig:II} for cross-sections), as well as all corresponding processes for the anti-deuteron. Finally, elastic $\pi d$, $N d$ and $\bar{N} d$ cross-sections were implemented to study the deuteron freeze-out. Their cross-sections are given in Tab. \ref{Tab:II}. All the details of the implementation, in particular the cross-sections, matrix elements and the detailed balance, are discussed in the Appendices \ref{AppendixA} and \ref{AppendixB}.

\section{Results} \label{Sec:Results}

\subsection{Transverse momentum spectra}

\begin{figure}[htp]
\includegraphics[width=0.5\textwidth]{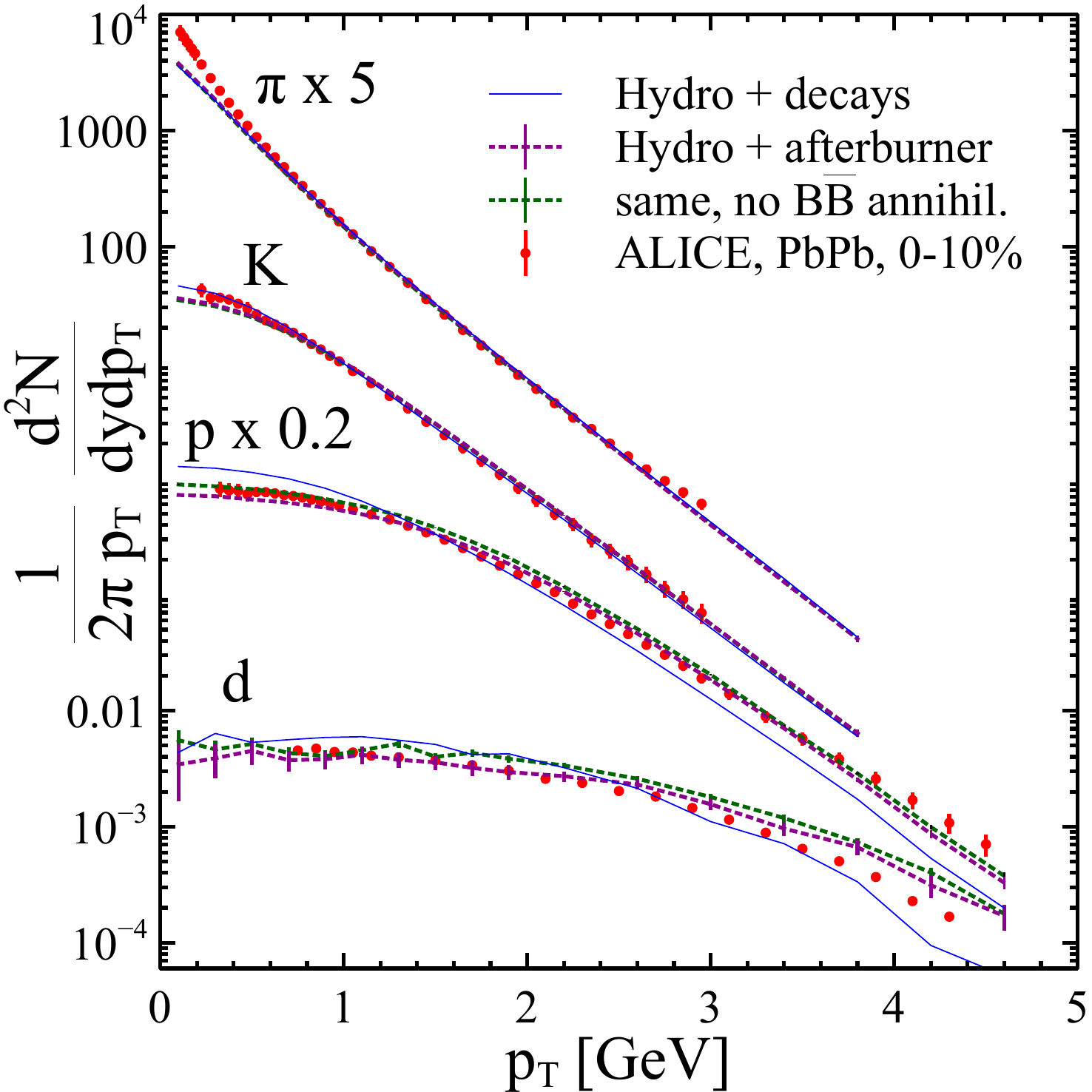}
\caption{Identified particle transverse momentum spectra compared to experimental measurements \cite{Adam:2015vda,Abelev:2012wca}.}
\label{Fig:III}
\end{figure}

The aforementioned implementation of deuteron production is now applied as a part of the afterburner in Pb+Pb collisions at $\sqrt{s_{NN}} = 2.76$ TeV. Our goal is not to fit experimental data as precisely as possible, but rather to understand the process of deuteron production qualitatively. Nevertheless, we first ensure that the transverse momentum spectra are reproduced reasonably well, see Fig \ref{Fig:III}. Pion and kaon spectra are described rather well already by hydrodynamics without afterburner, and the effect of the rescatterings in the afterburner does not exceed a few percent. This is different for nucleons: the $p_T$-spectrum produced by hydrodynamics and subsequent resonance decays is significantly modified by the afterburner. Large part of this modification is the effect of the so-called pion wind: protons rescatter with pions gaining higher transverse momentum. As shown in Fig. \ref{Fig:III}, $B\bar{B} \to $ mesons annihilations also influence the $p_T$ spectra, decreasing the number of protons both at small and large $p_T$ by about 10\%. The thermal deuteron yield produced on the hydrodynamic hypersurface is close to the experimentally measured, but the $p_T$ spectrum overshoots at low $p_T$ and undershoots at high $p_T$. The agreement, especially at low $p_T$, is improved by the afterburner.

\subsection{Parameter-free calculation of deuteron $B_2(p_T)$}

From the transverse momentum spectra shown in Fig. \ref{Fig:III} we obtain the coalescence parameter $B_2(p_T)$ by dividing the final spectra from Fig. \ref{Fig:III} in dashed dark-magenta line (color online):

\begin{eqnarray}
 B_2 (p_T) = \frac{\frac{1}{2\pi} \frac{d^3 N_d}{p_T dp_T dy} |_{p_T^d = 2 p_T^p}}{\left(\frac{1}{2\pi} \frac{d^3 N_p}{p_T dp_T dy} \right)^2}
\end{eqnarray}

\begin{figure}[htp]
\includegraphics[width=0.5\textwidth]{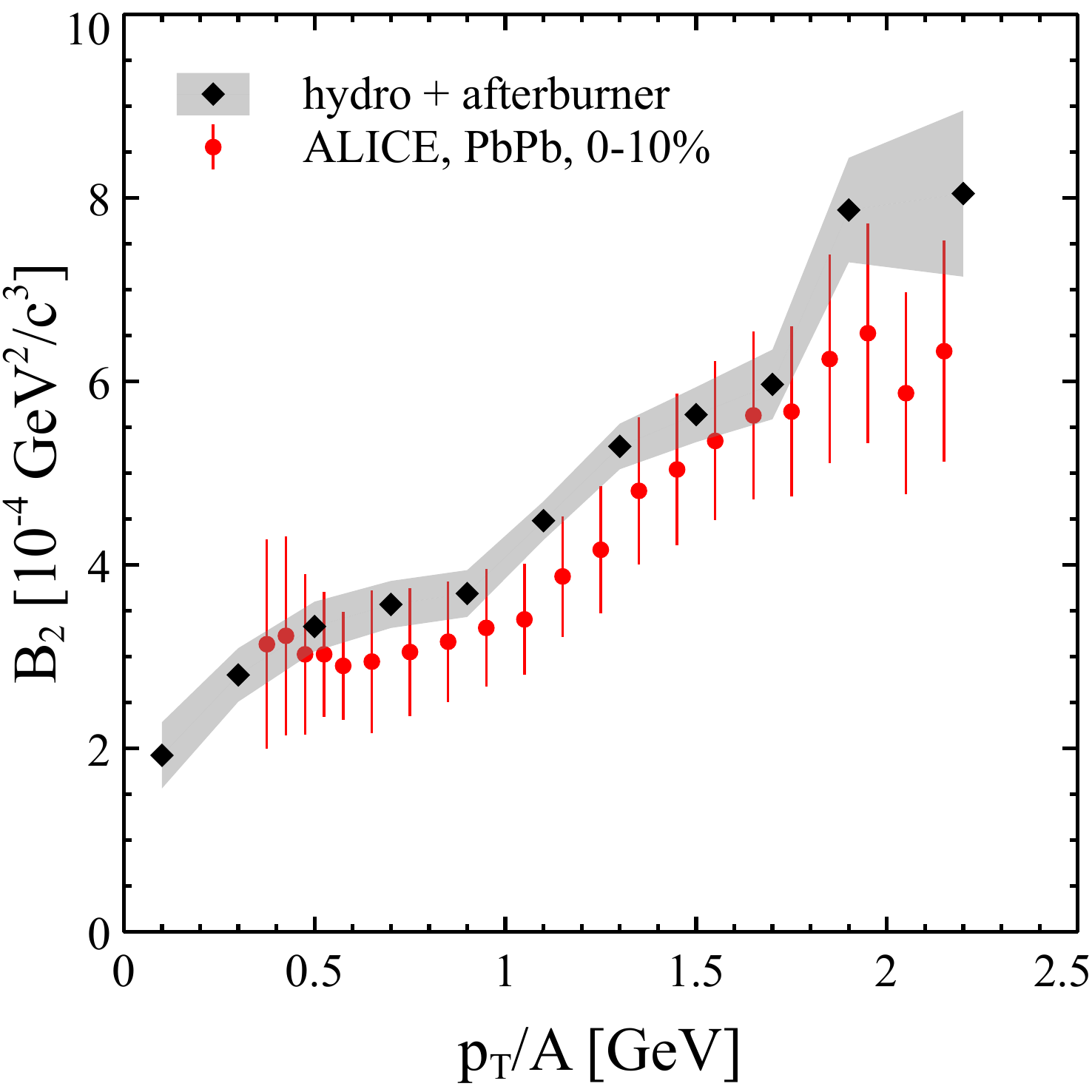}
\caption{Deuteron coalescence parameter $B_2$, extracted from the hydro + SMASH simulation (which does not involve any coalescence, only collisions with experimentally known deuteron cross-sections) is compared to ALICE measurement \cite{Acharya:2017dmc}.}
\label{Fig:IV}
\end{figure}

As one can see in Fig. \ref{Fig:IV}, our simulation reproduces the measured dependence of the $B_2$ coalescence parameter on transverse momentum rather well. In Fig. \ref{Fig:III} one can see that both proton and deuteron spectra are slightly overestimated. Since $B_2$ is reproduced well, we conclude that it is the over-prediction of the proton spectrum at $p_T = 1-2.5$ GeV that leads to the overestimation of the deuteron spectrum at $2 p_T = 2 - 5$ GeV. Therefore, an improved result for the proton spectrum should lead to a better description of the deuteron spectrum. We note, that a similar calculation of $B_2$ using the AMPT transport model overestimates $B_2$ by at least factor 2 \cite{Acharya:2017dmc}. The main deuteron production mechanism in the hadronic part of AMPT in this calculation appears to be the  $d\pi \leftrightarrow NN$ reactions. These reactions have small cross-section, as we show in Fig. \ref{Fig:I}, and therefore we would rather expect considerable underestimation of $B_2$ by AMPT. 

Unlike many coalescence models, such as \cite{Sombun:2018yqh}, our model does not require any tunable parameters to compute $B_2(p_T)$. It is in principle possible to compute $B_2(p_T)$ in the advanced version of coalescence model without tunable parameters~\cite{Danielewicz:1992pei,Scheibl:1998tk}. Although the methodology is available, to our best knowledge, so far it has not been applied to heavy ion collisions.

\subsection{Deuteron evolution in the afterburner stage}

\subsubsection{Chemical and kinetic freeze-out}
As we discussed above, the deuteron transverse momentum spectrum has changed during the afterburner evolution. However, the deuteron midrapidity yield turns out to be almost constant in time. Does it mean that deuteron has undergone chemical freeze-out at $T = 155$ MeV, as thermal model assumes? The answer turns out to be no: less than 1\% of  the final deuterons originate from hydrodynamics directly and are not destroyed in the afterburner.

\begin{figure}[htp]
\includegraphics[width=0.5\textwidth]{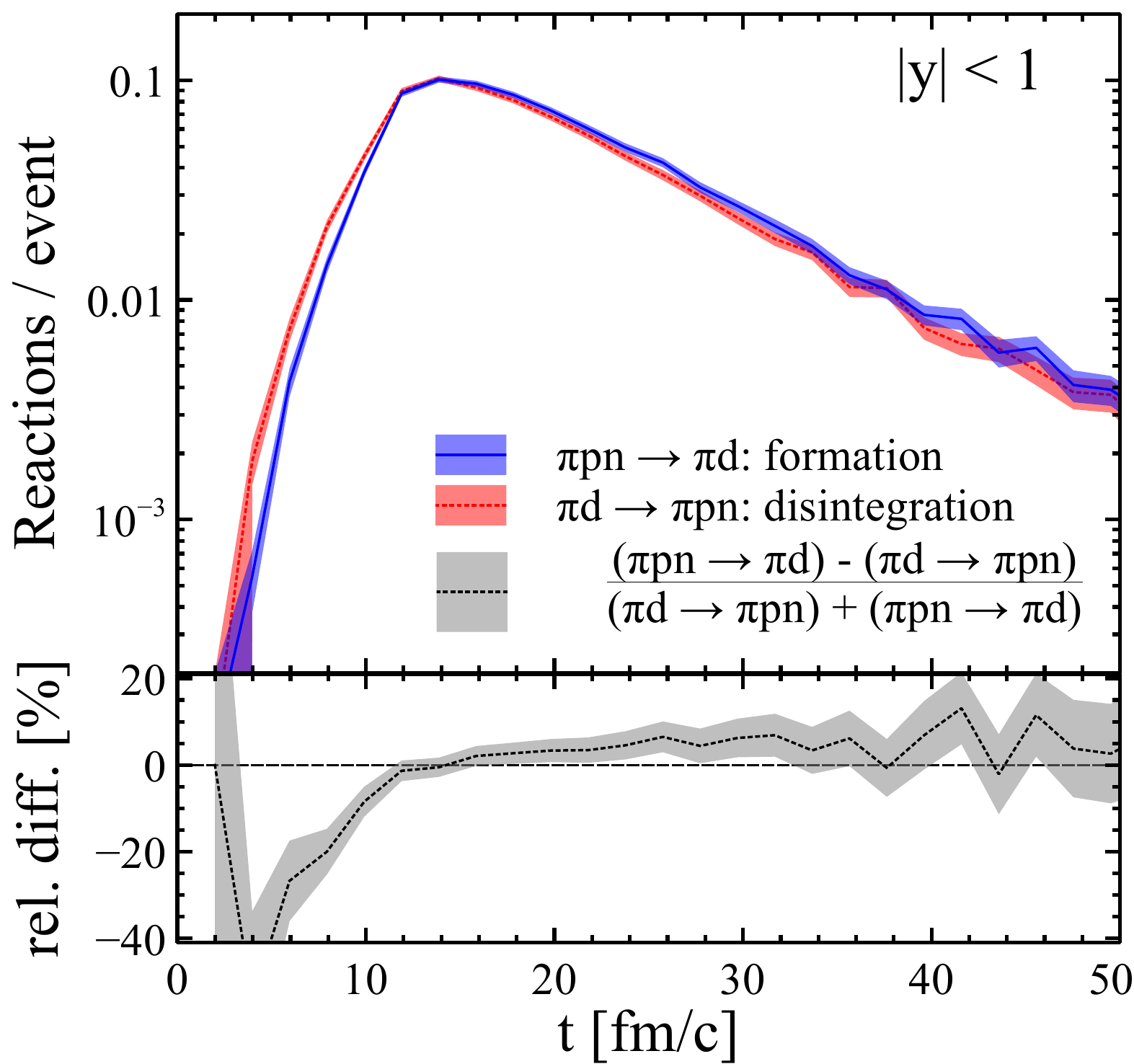}
\caption{Reaction rates of the most important $\pi d \leftrightarrow \pi p n$ reaction in forward and reverse direction.}
\label{Fig:V}
\end{figure}

Constant destruction and creation of deuterons, mainly in $d\pi \leftrightarrow np\pi$ reactions, occurs in the afterburner. This statement has to be understood in a statistical sense, because there is only one deuteron per unit of rapidity per 10 events produced. One can ask, if the  $d\pi \leftrightarrow np\pi$ reaction reaches equilibrium in the statistical sense. The answer is provided by Fig. \ref{Fig:V}: starting from $t \simeq 10-15$ fm/c the average number of forward and reverse reactions are close. At the earlier time the disintegration of deuterons, which were created on the hydrodynamic hypersurface, dominates. Later, formation exceeds disintegration by around 5\%. The difference is small, but it is statistically significant. Other inelastic reactions involving deuterons, such as $Nd \leftrightarrow N np$ and $\bar{N}d \leftrightarrow \bar{N} np$, are significantly less equilibrated, mostly favoring deuteron disintegration. However, the total number of those reactions is around 10 times smaller than $d\pi \leftrightarrow np\pi$.

\begin{figure}[htp]
\includegraphics[width=0.5\textwidth]{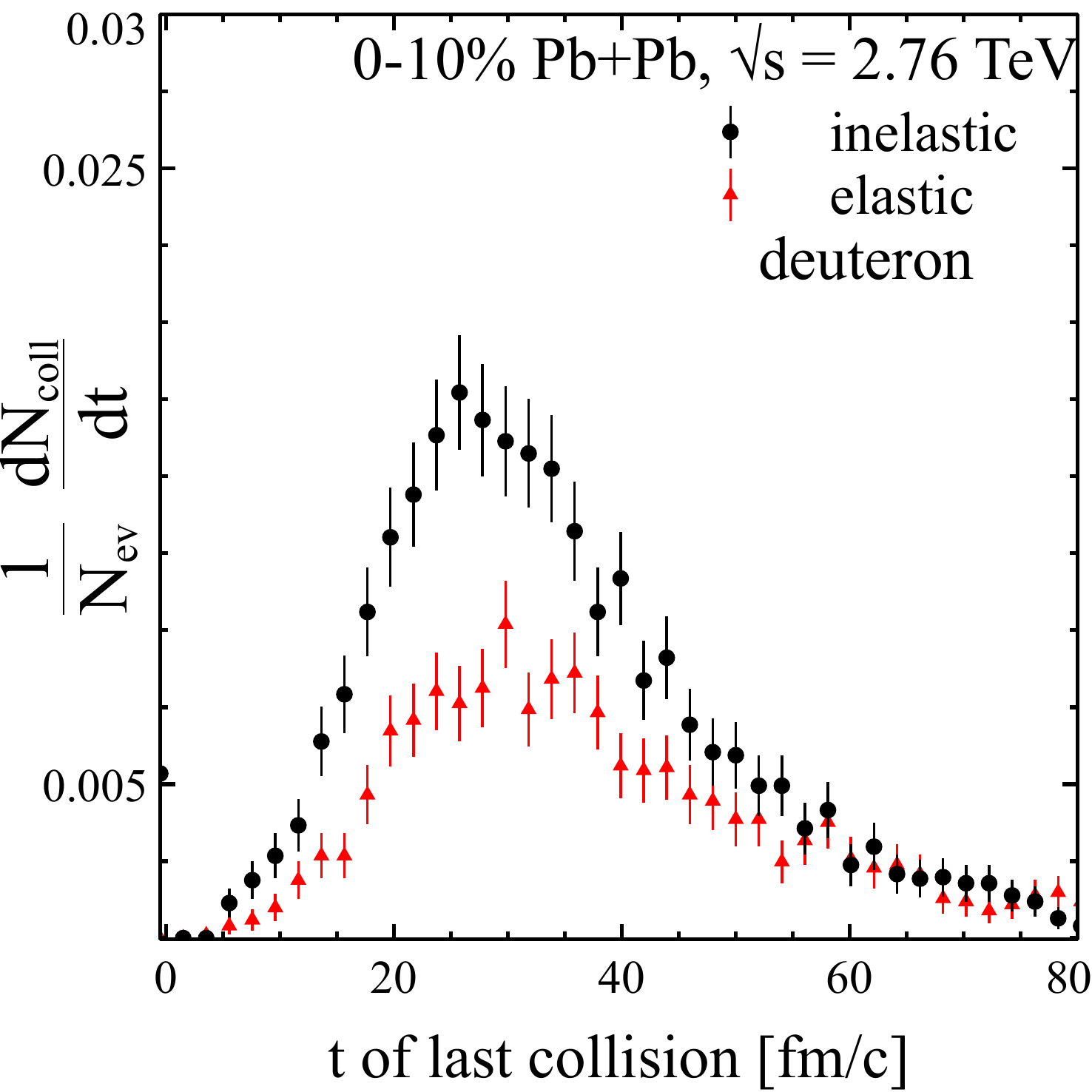}
\caption{Distribution of deuteron last collision time in case last collision was inelastic (circles) or elastic (triangles). Kinetic freeze-out for deuterons coincides with chemical freeze-out, which follows from inelastic $\pi d$ cross-section being larger than elastic, see Fig. \ref{Fig:I}.}
\label{Fig:VI}
\end{figure}

The chemical freeze-out of deuterons occurs as late as 25-35 fm/c, as can be seen in  Fig. \ref{Fig:VI}, where we show the distribution of the last collision times for both elastic and inelastic collisions. Also, both the elastic and inelastic collision time distributions peak around the same time indicating that kinetic and chemical freeze-out more or less coincide. This is in contrast to a usual freeze-out scheme, common for most of the hadrons --- first chemical freeze-out, then thermal. The reason is the following. For most of the hadrons inelastic cross-sections are smaller than elastic. Therefore during the expansion inelastic collisions cease first (chemical freeze-out), but elastic still continue until the kinetic freeze-out. The deuteron-pion elastic cross-section, on the other hand, is smaller than inelastic (see Fig. \ref{Fig:I}), that is why chemical and kinetic freeze-out of deuterons are approximately simultaneous.

\subsubsection{Deuteron yield} \label{Sec:deuteron_yield}

We have shown above that the deuteron chemical freeze-out does not occur at temperature $T=155$ MeV, but rather at the later stage, when the temperature is lower. However, our final deuteron yield is very close to the chemically equilibrated yield at $T=155$ MeV. A priori there is no obvious reason to expect that $\pi d \leftrightarrow \pi n p$ reactions close to equilibrium, together with $B\bar{B}$ annihilations {\it out of equilibrium}\footnote{While the cross-section at the most likely $\sqrt{s}$ of $B\bar{B}$ collisions is significant: 50-80 mb, the density of baryons and hence the reaction rate is small so the $B\bar{B}$ processes quickly fall out of equilibrium}, 
should not change the deuteron yield. Is it a mere coincidence or is there an underlying  physical reason?  
To understand this we consider five scenarios:

\begin{enumerate}
\item The default scenario: the number of deuterons sampled at particlization is in full chemical equilibrium with the hadrons. We denote the event average in this case as $N_{d}^{th}$. This is the scenario used in the previous sections.
\item Initial excess of deuterons: we artificially sample 3 times more deuterons at particlization than in the default scenario.
\item No deuterons from hydrodynamics. All the deuterons are produced in the afterburner.
\item No $B\bar{B}$ annihilations: same as default, but $B\bar{B}$ annihilations in the afterburner are switched off. This allows to assess the role of annihilations.
\item Same as the default scenario except that we perform the  particlization at $T = 165$ MeV instead of $T = 155$ MeV.
\end{enumerate}

\begin{figure}[htp]
\includegraphics[width=0.5\textwidth]{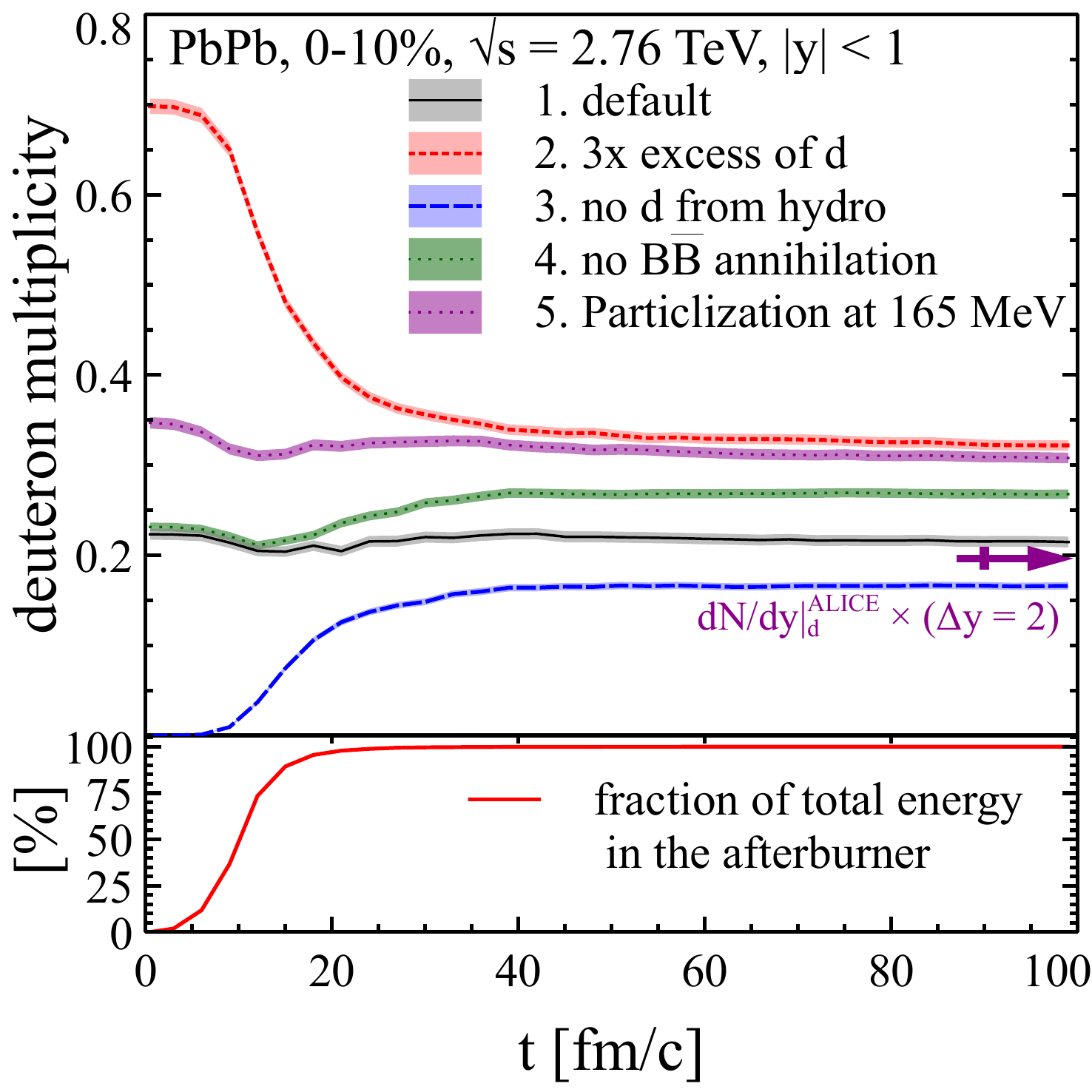}
\caption{Upper panel: deuteron yield (both in hydrodynamics and afterburner) versus time for the scenarios, described in the text. Lower panel: relative amount of energy in the afterburner. This is to indicate, how much of the system is already treated by the afterburner.}
\label{Fig:mult_vs_time}
\end{figure}

In the first three scenarios the only difference is the sampled amount of deuterons from the hydrodynamics. In Fig. \ref{Fig:mult_vs_time} one can see that in these three cases the $\pi d \leftrightarrow \pi n p$ reactions have almost enough time to drive deuteron yield to the same value, defined by the average phase space density of the nucleons in the system. Besides this, the scenario, where no deuterons are produced from hydrodynamics, is interesting by itself, because it still leads to a deuteron yield comparable to the one experimentally measured. First of all, this demonstrates how large the effect of the afterburner is for deuteron production. Indeed, all the deuterons in this scenario are produced from the afterburner. Second, it allows to conjecture that in the actual experiment all deuterons are produced in the hadronic phase. Our calculation does not prove it, but shows that such a scenario is possible.

The fourth scenario in Fig. \ref{Fig:mult_vs_time} shows the role of $B\bar{B}$ annihilations. Without annihilations the final deuteron yield is 20\% larger. If the detailed balance for the $B\bar{B} \leftrightarrow$ mesons processes was fulfilled, in other words if the processes mesons $\to B\bar{B}$ were implemented in SMASH, the final deuteron yield would be between our default calculation and the scenario without annihilation. This also seems to suggest that the unchanged deuteron yield in time is a coincidence. In the fifth scenario we show, however, that this coincidence persists if one changes the temperature of particlization to 165 MeV. This is surprising. To understand this better we proceed with a toy model of deuteron production, that reproduces our results qualitatively and explains the coincidence.

\subsubsection{Toy model of deuteron production} \label{Sec:toy_model}

From the previous section and Fig. \ref{Fig:mult_vs_time} we conclude that $\pi d \leftrightarrow \pi np$ reactions close to equilibration tend to increase the amount of deuterons, if annihilations are absent. At the same time, $B\bar{B}$ annihilations out of equilibrium decrease the amount of nucleons and consequently, the amount of deuterons. As a result, when both mechanisms are at play, the deuteron yield stays approximately constant. This balance of two mechanisms, however, turns out to be surprisingly stable with respect to the change of particlization temperature. To explain it in an intuitive way, let us consider a simple thermodynamic toy-model, which assumes isentropic expansion, constant number of pions and complete absence of $B\bar{B}$ annihilations. For simplicity, we consider an expanding ideal gas of pions, nucleons, deltas, and deuterons. The corresponding equations are:
\begin{align}
&(\rho_{\Delta}(T, \mu_B + \mu_{\pi}) + \rho_{\pi}(T, \mu_{\pi})) V =& const   \\
&(\rho_N(T, \mu_B) + \rho_{\Delta}(T, \mu_B + \mu_{\pi}) +\nonumber &\\ &+ 2 \rho_d(T, 2 \mu_B)) V =& const  \\
&(s_{\pi}(T, \mu_{\pi}) + s_N(T, \mu_B) +&\nonumber \\
&+ s_{\Delta}(T, \mu_B + \mu_{\pi}) + s_d(T, 2 \mu_B)) V =& const
\end{align}

Here $\rho$ is the density and $s$ is the entropy density computed as

\begin{eqnarray}
\rho = \frac{g}{2 \pi^2 \hbar^3} e^{\mu/T} m^2 T K_2(m/T) \\
s = \frac{dp}{dT} = \rho + T \frac{\partial\rho}{\partial T}
\end{eqnarray}

\begin{figure}
\includegraphics[width=0.49\textwidth]{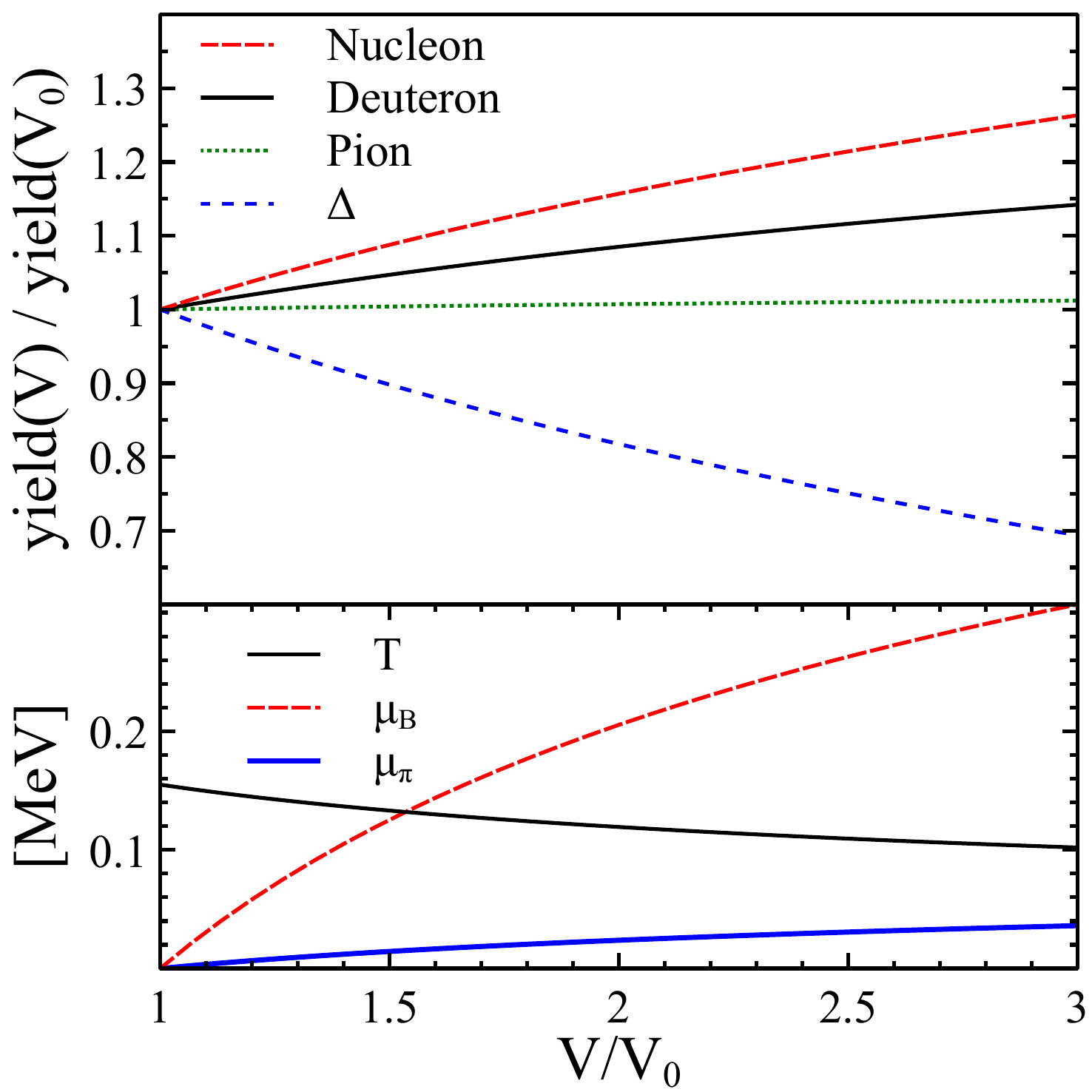}
\caption{Evolution of yields (upper panel) and thermodynamic variables (lower panel) in our toy model without annihilations for $T_0 = 155$ MeV. The deuteron yield grows, which is similar to our simulation within the fourth scenario in Fig. \ref{Fig:mult_vs_time}.}
\label{Fig:VIII}
\end{figure}

Here $g$ is degeneracy of a particle, $m$ is its mass, and $\mu$ is its chemical potential. We assume that initially the system has temperature $T_0$, volume $V_0$, $\mu_B = 0$ and $\mu_{\pi} = 0$. As the system expands and the volume increases, the temperature drops and the fugacity parameters, $\mu_B$ and $\mu_{\pi}$, increase. In Fig.\ref{Fig:VIII} one can see that the deuteron yield indeed grows, in qualitative agreement with our simulation without annihilations in Fig. \ref{Fig:mult_vs_time}. The picture remains almost unchanged, when $T_0$ is set to 165 MeV instead of 155 MeV. To emulate annihilations, that are out of equilibrium, and  quickly freeze-out, in the solution of our toy-model equation we set

\begin{eqnarray}
\mu_B \to \mu_B \frac{V/V_0}{a + V/V_0}\,,
\end{eqnarray}

so that at large $V/V_0$ annihilations are not acting, while at smaller $V/V_0$ they are the strongest. Parameter $a$ regulates the strength of annihilations: if it is large, then $\mu_B \to 0$, which corresponds to very effective annihilations; if $a = 0$ then it is our initial model without annihilations. Setting $a = 0.1$ we obtain relative deuteron yield, which first decreases and then increases by about 5\%. Setting $T_0$ to 165 MeV instead of 155 MeV leaves this picture almost unchanged. This is in qualitative agreement with our simulation and explains the persistent coincidence between final deuteron yield and thermal yield at particlization temperature.

\section{Summary and Discussion} \label{Sec:Discussion}

We have studied deuteron production in heavy ion collisions at LHC energies using a hybrid (hydrodynamics + hadronic afterburner) approach. We simulate deuterons in the afterburner as pointlike particles with experimentally known interaction cross-sections. Our investigation provides the following picture of deuteron production in heavy ion collisions at LHC energies. Whether deuterons are created during the hydrodynamic stage or not, in the hadronic afterburner stage the reactions $\pi d \leftrightarrow \pi pn$ drive deuterons close to statistical equilibrium (Fig. \ref{Fig:mult_vs_time}). Even though the cross-section of $\pi d \leftrightarrow \pi pn$ is large (Fig. \ref{Fig:I}), there is not sufficient time to equilibrate deuterons completely, because the system expands and reactions eventually freeze-out. We further find that chemical and kinetic freeze-outs of deuterons roughly coincide (Fig. \ref{Fig:VI}), because the inelastic cross-sections of deuterons with pions is larger than the elastic one, unlike for most of the hadrons. We also observe that baryon-antibaryon annihilation is important for deuteron production at high energies, because it changes the amount of nucleons, to which the amount of deuterons is related.

Clearly, deuterons do not freeze-out at $T = $155 MeV. In contrast, their yield can change until late times of the evolution. This is best demonstrated by the scenario with no deuterons at all at the end of the hydrodynamical stage, depicted by the blue line  in Fig. \ref{Fig:mult_vs_time}. One may still wonder, why the thermal model explains deuteron multiplicity well, even though its assumptions are violated. In Fig. \ref{Fig:mult_vs_time} one can see that the deuteron yield does not change much during the evolution, if initialized thermally, {\it as if} a chemical freeze-out would occur. We explain this by an interplay of $\pi d \leftrightarrow \pi pn$ reactions, which are close to equilibration, and $B\bar{B}$ annihilations, that are far from equilibration. The first tend to increase the amount of deuterons above the thermal model value. The second reduce the amount of nucleons in the system, and, as a consequence, the amount of deuterons. With both acting together the amount of deuterons does not change much, if initialized thermally. It will be interesting to test this explanation at lower collision energies, where antibaryons are much less abundant and $B\bar{B}$ annihilations almost do not occur. If correct, we predict a larger deuteron yield than the thermal model. Here we limit ourselves to central collisions, but it turns out that our model works well for peripheral collisions too \cite{Oliinychenko:2018odl}.

In this study we have constrained ourselves to very high energy Pb+Pb collisions. However, we have developed an instrument that allows to study deuteron production not only at high energies, but also at intermediate ones, covered by RHIC Beam Energy Scan program. At these energies, the  reactions $\pi d \leftrightarrow \pi np$ are competing with $N d \leftrightarrow N np$. We conjecture that switching from one production process to another might be one reason behind the measured non-monotonic $B_2$ behavior against energy \cite{Yu:2017bxv}. Testing this conjecture will be a straightforward application of our results. It will also be extremely interesting to study the relation between the deuteron production and the nucleon density fluctuations at these energies. Can hadronic transport without a phase transition reproduce the measured bump in the collision energy dependence of $N_t N_p / N_d^2$ ratio \cite{Dingwei_Zhang}, which is claimed to be related to phase transition \cite{Sun:2018jhg}? We hope to answer this question in future publications.

\begin{acknowledgments}
D. O. thanks I. Karpenko for explaining the effects of hybrid approach parameters on the final hadron spectra; M. Ploskon for pointing to ALICE data; I. Strakovsky for pointing to deuteron cross sections and help with the SAID database; J. Schukraft, U. Heinz, and J. Aichelin for insightful discussions. D. O. and V. K. were supported by the U.S. Department of Energy, 
Office of Science, Office of Nuclear Physics, under contract number 
DE-AC02-05CH11231 and received support within the framework of the
Beam Energy Scan Theory (BEST) Topical Collaboration.
L.-G.P. was supported by the National Science Foundation (NSF) within the framework of the JETSCAPE collaboration, under grant number ACI-1550228. H. E. acknowledges funding of a Helmholtz Young Investigator Group VH-NG-822 from the Helmholtz Association and GSI. Computational resources were provided within the framework of the Landes-Offensive zur Entwicklung Wissenschaftlich-\"Okonomischer Exzellenz (LOEWE) program launched by the State of Hesse.

\end{acknowledgments}

\appendix
\section{Implementation of deuteron interactions}
\label{AppendixA}

The following reactions with deuterons were implemented in SMASH for this work: $\pi d \leftrightarrow NN$, $\pi d \leftrightarrow \pi np$, elastic $\pi d \to \pi d$, $N d \leftrightarrow N np$, and $\bar{N} d \leftrightarrow \bar{N} np$. The analogous set of reactions, only CPT conjugated, was implemented for anti-deuterons.

Many of the above mentioned reactions create deuterons via $3\to 2$ collisions, which are not measured in the experiment. Luckily, they can be connected to the measured reverse $2\to 3$ reactions using the detailed balance (or time reversal invariance) principle: $|M_{2\to 3}|^2 = |M_{3\to 2}|^2$, where $M_{2\to 3}$ and $M_{3\to 2}$ are the matrix elements of the corresponding reactions. Still, the $2\leftrightarrow 3$ processes, such as $d\pi \leftrightarrow np \pi$, $N d \leftrightarrow N n p$, and $\bar{N} d \leftrightarrow \bar{N} n p$, are challenging to implement within a transport approach with geometric collision treatment, which SMASH is. Implementing $2\leftrightarrow 3$ collisions, which obey the detailed balance principle, requires ``stochastic rates'' collision treatment, as applied e.g. in the BAMPS parton cascade for $gg\leftrightarrow ggg$ reaction \cite{Xu:2004mz}, or in other codes for $B\bar{B} \leftrightarrow 3M$ annihilation \cite{Pan:2014caa,Seifert:2017oyb}. We decided to avoid this rigorous, but technically involved approach. Instead we split $2 \leftrightarrow 3$ reactions into two $2\to 2$ reactions in the following way: $d \pi \leftrightarrow d'\pi \leftrightarrow np \pi$, $d N \leftrightarrow d'N \leftrightarrow np N$, and $d \bar{N} \leftrightarrow d'\bar{N} \leftrightarrow np \bar{N}$ via $d' \leftrightarrow np$. This requires introduction of a fictitious dibaryon resonance $d'$, whose mass and width are not fixed by the experimental data. Note that $d'$ is just a technical artifact of our method, different from the actual dibaryon resonance $d^*$ measured by WASA experiment \cite{Bashkanov:2015xsa}.

We set the pole mass $m_{d'} = m_{d} + 10$ MeV and width $\Gamma_{d'} = 100$ MeV, the spin of $d'$ is assumed to be 1. The motivation for this width is to have the $d'$ lifetime close to the time that proton and neutron spend flying past each other. Changing the width to 50 MeV or 200 MeV, together with refitting matrix elements to match experimental cross-sections, only changes the final deuteron yield within statistical errors. Changing the mass $m_{d'}$ changes results, but not much: larger $m_{d'}$ (again after refitting the matrix elements) causes a smaller final deuteron yield. For $m_{d'} = m_{d} + 100$ MeV the deuteron yield is around 15\% smaller than for $m_{d'} = m_{d} + 10$ MeV. This rather low sensitivity to variations of the parameters for $d'$ justifies our approach.

The $d'$ spectral function is computed in the same way as for any other resonance in SMASH:
\begin{eqnarray}
  \mathcal{A}(m) \, dm = \mathcal{N }\frac{2 m^2 \Gamma_{d'}(m)}{(m^2 - m_{d'}^2)^2 + m^2 \Gamma_{d'}^2(m)} \, dm \,,
\end{eqnarray}
where the factor $\mathcal{N}$ is chosen to fulfill  $\int_{m_{Th}}^{\infty} \mathcal{A}(m) dm = 1$. For further convenience, let us define the following functions:
\begin{eqnarray}
p_{cm}^2(\sqrt{s}, m_a, m_b) = \frac{(s + m_a^2 - m_b^2)^2}{4 s} - m_a^2 \\
F_{Xd \to Xd'} = \frac{1}{s}\int\displaylimits_{2 m_N}^{\sqrt{s} - m_X} \frac{p_{cm}(\sqrt{s}, m, m_X)}{p_{cm}(\sqrt{s}, m_{d}, m_X)} \,\mathcal{A}(m) \, dm \\
F_{Xd' \to Xd} = \frac{1}{s} \frac{p_{cm}(\sqrt{s}, m_{d}, m_X)}{p_{cm}(\sqrt{s}, m_{d'}, m_X)}
\end{eqnarray}

With their help the tables of matrix elements (Tab. \ref{Tab:I}) and cross-sections (Tab. \ref{Tab:II}) are written. The treatment of deuteron and $d'$ reactions is similar to other reactions with resonances in SMASH, for example $NN \leftrightarrow N\Delta$ or $K N \leftrightarrow K \Delta$, see \cite{Weil:2016zrk} for more details. This treatment fulfills the detailed balance relations. The angular distributions of the reaction products are assumed to be isotropic in the center of mass frame of the reaction.

\begin{table*}
\renewcommand{\arraystretch}{2.0}
\begin{tabular}{cc}
\toprule
      Reaction             &          Matrix element, $|M|^2$ \\
     $\pi d - NN$    &     $\displaystyle \frac{0.055}{(\sqrt{s} - 2.145)^2 + 0.065^2} \, (1 - \exp(- (\sqrt{s}- 2.0)/0.05))$    \\
     
     $\pi d - \pi d'$    &     $\displaystyle 295.5 + \frac{2.862}{0.53^2 + (\sqrt{s} - 2.181)^2} +
          \frac{0.0672}{(\sqrt{s} - m_{\pi} - 2 m_N)^2} - \frac{6.61753}{\sqrt{s} - m_{\pi} - 2 m_N} $   \\
     
     $N d - N d'$    &     $\displaystyle \frac{79.0474}{(\sqrt{s}-3 m_N)^{0.7897}} + 654.596 \, (\sqrt{s}-3 m_N)$    \\
     
     $\bar{N} d - \bar{N} d'$    &     $342.572 / (\sqrt{s}-3 m_N)^{0.6} $  \\
\botrule
\end{tabular}
\caption{Parametrized matrix elements in the reactions with deuterons. In some cases constant factors from the cross-section formulas are included into the matrix elements, see Tab. \ref{Tab:II}.}
\label{Tab:I}
\end{table*}

\begin{table*}
\renewcommand{\arraystretch}{1.6}
\begin{tabular}{lc}
\toprule
  Reaction    & Cross-section [mb] \\
  $np \to d'$ &  $\displaystyle \frac{3}{4} \frac{2\pi^2 (\hbar c)^2}{p^2_{cm}(\sqrt{s},m_N,m_N)} \mathcal{A}(\sqrt{s}) \Gamma_{d'}(\sqrt{s}) $  \\
  $\pi d \to \pi d $ (el)   &    $\displaystyle \frac{0.27}{(\sqrt{s} - 2.172)^2 + 0.065^2} + 4$  \\
  $N d \to N d$ (el)   &  $\displaystyle 10+ 2500 \, e^{- \frac{x^2}{0.003}} +  600 \, e^{- \frac{x^2}{0.1}}$, $x = s - 7.93$ \cite{Oh:2009gx} \\
  $\bar{N} d \to \bar{N} d$ (el) &    $= \sigma^{el}_{N d \to N d}$   \\
  $\pi d \to \pi d'$ &  $\displaystyle 9|M_{\pi d - \pi d'}|^2  F_{\pi d \to \pi d'} $  \\
  $\pi d' \to \pi d$ &  $\displaystyle 9|M_{\pi d - \pi d'}|^2  F_{\pi d' \to \pi d} $    \\
  $N d \to N d'$ &    $\displaystyle 6|M_{N d - N d'}|^2 F_{N d \to N d'}$   \\
  $N d' \to N d$ &    $\displaystyle 6|M_{N d - N d'}|^2 F_{N d' \to N d}$   \\
  $\bar{N} d \to \bar{N} d'$ &    $\displaystyle 6|M_{\bar{N} d - \bar{N} d'}|^2 F_{\bar{N} d \to \bar{N} d'} $   \\
  $\bar{N} d' \to \bar{N} d$ &    $\displaystyle 6|M_{\bar{N} d - \bar{N} d'}|^2 F_{\bar{N} d' \to \bar{N} d}$   \\
  $\pi^+ d \to pp$, $\pi^- d \to nn$   &     $\displaystyle \frac{2|M_{\pi d - N N}|^2}{s} \frac{p_{CM}(\sqrt{s}, m_N, m_N)}{p_{CM}(\sqrt{s}, m_d, m_{\pi})}$          \\
  $\pi^0 d \to np$   &     $= \frac{1}{2} \sigma_{\pi^+ d \to pp}$          \\
  $NN \to \pi d$   &            $\displaystyle \frac{2|M_{\pi d - N N}|^2}{s} \frac{p_{CM}(\sqrt{s}, m_d, m_{\pi})}{p_{CM}(\sqrt{s}, m_N, m_N)}$  \\
\botrule
\end{tabular}
\caption{Deuteron cross-sections: either parametrizations or their relations to parametrized matrix elements from Tab. \ref{Tab:I}, with the account of the spin and symmetry factors.}
\label{Tab:II}
\end{table*}

\section{Testing the detailed balance} \label{AppendixB}

\begin{figure}
  \includegraphics[width=0.5\textwidth]{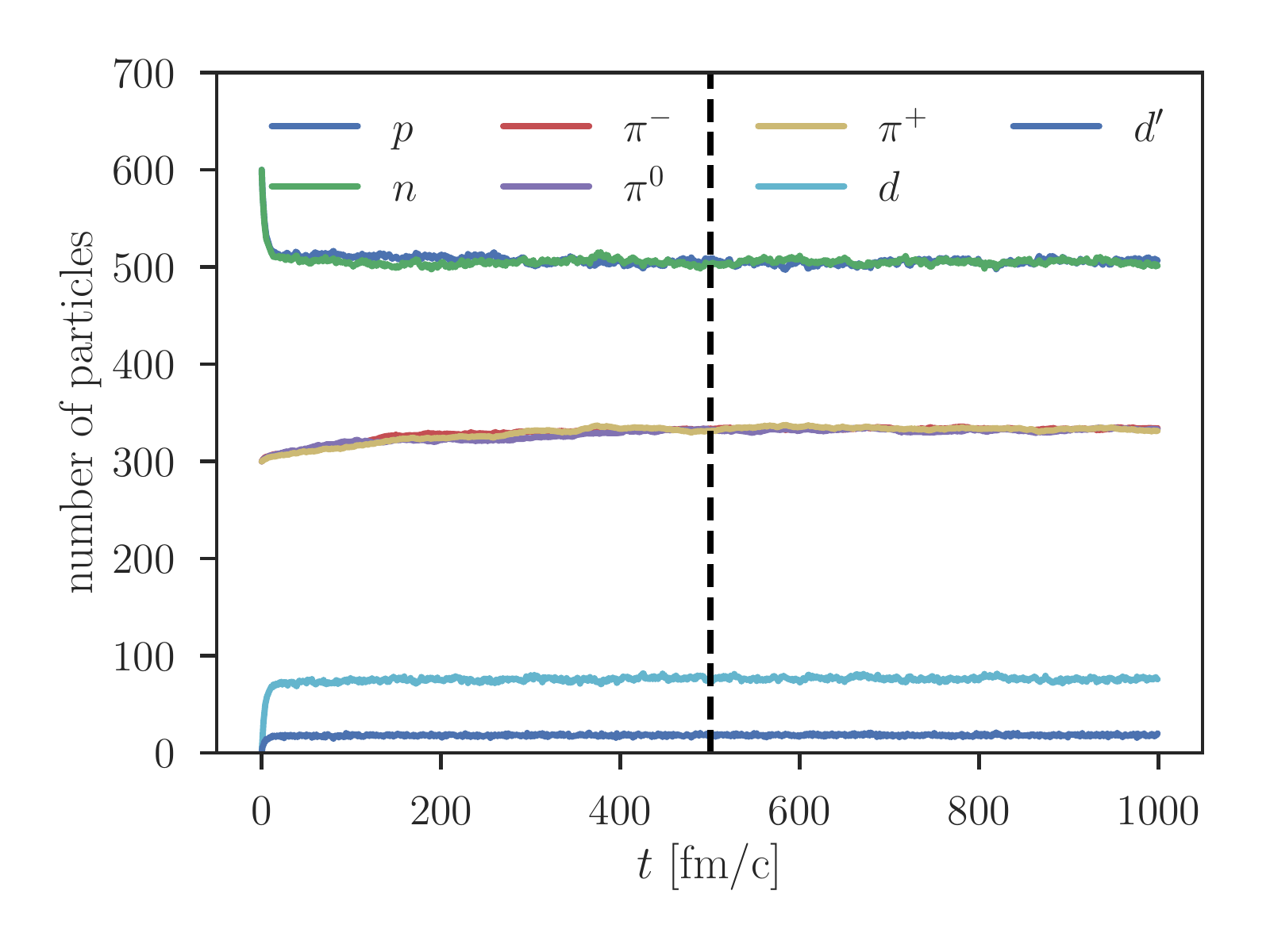}
  \caption{Equilibration in the box, where interactions are artificially limited to those involving deuterons. Nucleon and deuteron numbers equilibrate for 10-15 fm/c via $\pi d \leftrightarrow \pi np$ and $N d \leftrightarrow N np$ reactions with large cross-sections. Pions equilibrate via $\pi d \leftrightarrow NN$, which has a small cross-section and takes around 500 fm/c to equilibrate.}
  \label{Fig:detbal_I}
\end{figure}

The principle of the detailed balance is at the core of this work, because it allowed to implement deuteron production reactions, that were not experimentally measured. This motivated us to test, how well the detailed balance is fulfilled in our implementation. Moreover, this test also provides some insights into the process of the chemical equilibration of deuterons.

For the test we initialize a (10 fm)$^3$ box with periodic boundary conditions with 60 protons, 60 neutrons and 30 of each of pion species. The initial distribution in coordinate space is uniform, in momentum space a thermal Boltzmann distribution with temperature $T = 155$ MeV has been applied. We use the testparticles ansatz, where the initial numbers of particles is multiplied by $N_{test}$, while all the cross-sections are divided by $N_{test}$, where $N_{test} = 10$. This helps to avoid spurious 3-body correlation effects and improves the detailed balance. For the test only reactions involving deuterons are allowed. As expected, nucleons and deuterons equilibrate fast, after around 10-15 fm/c, while pions need around 500 fm/c, see Fig. \ref{Fig:detbal_I}. This is because the $\pi d \leftrightarrow \pi np$ and $N d \leftrightarrow N np$ cross-sections are much larger than $\pi d \leftrightarrow NN$ --- the only reaction that changes the pion number in our setup.

\begin{figure*}
  \includegraphics[width=\textwidth]{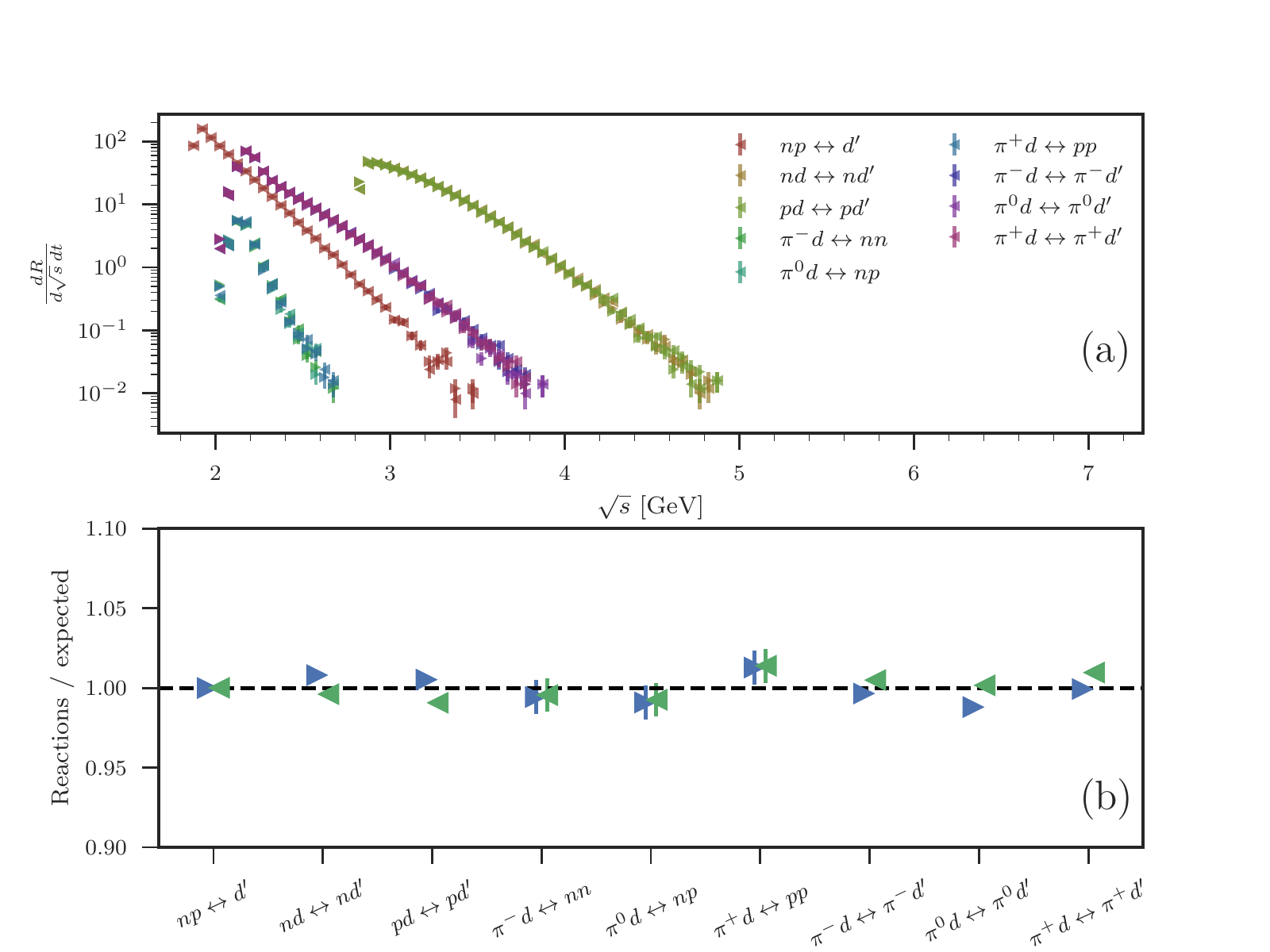}
  \caption{Demonstration, that detailed balance is obeyed with 1\% precision. Differential reaction rates against the center of mass energy of the reaction (a) and integrated numbers of reactions, scaled over isospin average (b) are shown. Forward reactions are marked by triangles pointing right, reverse reactions -- by triangles pointing left. In case of a perfect detailed balance both should point at the same value.}
  \label{Fig:detbal_II}
\end{figure*}

As soon as the system exhibits apparent chemical equilibration, we check, if the numbers of forward and reverse reactions coincide, both for total numbers (Fig. \ref{Fig:detbal_II}b) and differential in the collision energy (Fig. \ref{Fig:detbal_II}a). In both cases they are equal within 1\%. The deviations from perfect detailed balance are mainly caused by the large cross-sections of $\pi d'$ and $N d'$, especially near the threshold, where the cross-sections diverge. The cutoff in the collision finding algorithm does not allow all the collisions to be found: this is the reason of the small detailed balance violations. Violations become considerably larger, up to 5\%, if many anti-nucleons are added to the system. The reason for these deviations is the similar --- large $\bar{N} d'$ cross-sections. This test allows us to conclude that the detailed balance in our deuteron implementation is obeyed with a good precision and the deviations from it are well-understood. To fulfill this test with high precision we have increased the usual 200 mb SMASH cross-section cutoff to 2000 mb. The 2000 mb cutoff was also used for the simulations throughout the article.


\begin{thebibliography}{99}

\bibitem{Adam:2015vda} 
  J.~Adam {\it et al.} [ALICE Collaboration],
  Phys.\ Rev.\ C {\bf 93}, no. 2, 024917 (2016)
  doi:10.1103/PhysRevC.93.024917
  [arXiv:1506.08951 [nucl-ex]].
  
\bibitem{Cern_courier}
  P.~Braun-M\"{u}nzinger, B. D\"{o}nigus and N. L\"{o}her,
  CERN Courier, August 2015.
  
\bibitem{Kapusta:1980zz} 
  J.~I.~Kapusta,
  Phys.\ Rev.\ C {\bf 21}, 1301 (1980).
  doi:10.1103/PhysRevC.21.1301
  
\bibitem{Siemens:1979dz} 
  P.~J.~Siemens and J.~I.~Kapusta,
  Phys.\ Rev.\ Lett.\  {\bf 43}, 1486 (1979).
  doi:10.1103/PhysRevLett.43.1486
  
\bibitem{Andronic:2010qu} 
  A.~Andronic, P.~Braun-Munzinger, J.~Stachel and H.~St\"{o}cker,
  Phys.\ Lett.\ B {\bf 697}, 203 (2011)
  doi:10.1016/j.physletb.2011.01.053
  [arXiv:1010.2995 [nucl-th]].
  
\bibitem{Andronic:2012dm} 
  A.~Andronic, P.~Braun-Munzinger, K.~Redlich and J.~Stachel,
  Nucl.\ Phys.\ A {\bf 904-905}, 535c (2013)
  doi:10.1016/j.nuclphysa.2013.02.070
  [arXiv:1210.7724 [nucl-th]].

  
\bibitem{Cleymans:2011pe} 
  J.~Cleymans, S.~Kabana, I.~Kraus, H.~Oeschler, K.~Redlich and N.~Sharma,
  Phys.\ Rev.\ C {\bf 84}, 054916 (2011)
  doi:10.1103/PhysRevC.84.054916
  [arXiv:1105.3719 [hep-ph]].
  
\bibitem{Oliinychenko:2016dtb} 
  D.~R.~Oliinychenko, K.~A.~Bugaev, V.~V.~Sagun, A.~I.~Ivanytskyi, I.~P.~Yakimenko, E.~G.~Nikonov, A.~V.~Taranenko and G.~M.~Zinovjev,
  arXiv:1611.07349 [nucl-th].
  

\bibitem{Sato:1981ez} 
  H.~Sato and K.~Yazaki,
  Phys.\ Lett.\  {\bf 98B}, 153 (1981).
  doi:10.1016/0370-2693(81)90976-X
  
\bibitem{Gutbrod:1988gt} 
  H.~H.~Gutbrod, A.~Sandoval, P.~J.~Johansen, A.~M.~Poskanzer, J.~Gosset, W.~G.~Meyer, G.~D.~Westfall and R.~Stock,
  Phys.\ Rev.\ Lett.\  {\bf 37}, 667 (1976).
  doi:10.1103/PhysRevLett.37.667

\bibitem{Mrowczynski:1992gc} 
  S.~Mrowczynski,
  Phys.\ Lett.\ B {\bf 277}, 43 (1992).
  doi:10.1016/0370-2693(92)90954-3
  
\bibitem{Csernai:1986qf} 
  L.~P.~Csernai and J.~I.~Kapusta,
  Phys.\ Rept.\  {\bf 131}, 223 (1986).
  doi:10.1016/0370-1573(86)90031-1

\bibitem{Mrowczynski:2016xqm} 
  S.~Mrowczynski,
  Acta Phys.\ Polon.\ B {\bf 48}, 707 (2017)
  doi:10.5506/APhysPolB.48.707
  [arXiv:1607.02267 [nucl-th]].
  
\bibitem{Bazak:2018hgl} 
  S.~Bazak and S.~Mrowczynski,
  arXiv:1802.08212 [nucl-th].

\bibitem{Dong:2018cye} 
  Z.~J.~Dong, G.~Chen, Q.~Y.~Wang, Z.~L.~She, Y.~L.~Yan, F.~X.~Liu, D.~M.~Zhou and B.~H.~Sa,
  arXiv:1803.01547 [nucl-th].
  
\bibitem{Sun:2016rev} 
  K.~J.~Sun and L.~W.~Chen,
  Phys.\ Rev.\ C {\bf 94}, no. 6, 064908 (2016)
  doi:10.1103/PhysRevC.94.064908
  [arXiv:1607.04037 [nucl-th]].
  
\bibitem{Sun:2018jhg} 
  K.~J.~Sun, L.~W.~Chen, C.~M.~Ko, J.~Pu and Z.~Xu,
  Phys.\ Lett.\ B {\bf 781}, 499 (2018)
  doi:10.1016/j.physletb.2018.04.035
  [arXiv:1801.09382 [nucl-th]].

\bibitem{Polleri:1997bp} 
  A.~Polleri, J.~P.~Bondorf and I.~N.~Mishustin,
  Phys.\ Lett.\ B {\bf 419}, 19 (1998)
  doi:10.1016/S0370-2693(97)01455-X
  [nucl-th/9711011].
  
\bibitem{Nagamiya:1981sd} 
  S.~Nagamiya, M.~C.~Lemaire, E.~M\"{o}ller, S.~Schnetzer, G.~Shapiro, H.~Steiner and I.~Tanihata,
  Phys.\ Rev.\ C {\bf 24}, 971 (1981).
  doi:10.1103/PhysRevC.24.971
  
\bibitem{Biro:1981es} 
  T.~Biro, H.~W.~Barz, B.~Lukacs and J.~Zimanyi,
  Phys.\ Rev.\ C {\bf 27}, 2695 (1983).
  doi:10.1103/PhysRevC.27.2695
  
\bibitem{Stoecker:1981kf} 
  H.~St\"{o}cker,
  Lawrence Berkeley Lab. - LBL-12302 (81,REC.JUL.) 12p
  
\bibitem{Csernai:1987ri} 
  L.~P.~Csernai, J.~I.~Kapusta, G.~I.~Fai, D.~Hahn, J.~Randrup and H.~St\"{o}cker,
  Phys.\ Rev.\ C {\bf 35}, 1297 (1987).
  doi:10.1103/PhysRevC.35.1297
  
\bibitem{Hirano:2012kj} 
  T.~Hirano, P.~Huovinen, K.~Murase and Y.~Nara,
  Prog.\ Part.\ Nucl.\ Phys.\  {\bf 70}, 108 (2013).
  
\bibitem{Petersen:2008dd} 
  H.~Petersen, J.~Steinheimer, G.~Burau, M.~Bleicher and H.~St\"{o}cker,
  Phys.\ Rev.\ C {\bf 78}, 044901 (2008)

\bibitem{Werner:2010aa} 
  K.~Werner, I.~Karpenko, T.~Pierog, M.~Bleicher and K.~Mikhailov,
  Phys.\ Rev.\ C {\bf 82}, 044904 (2010)
  
\bibitem{Ryu:2012at} 
  S.~Ryu, S.~Jeon, C.~Gale, B.~Schenke and C.~Young,
  Nucl.\ Phys.\ A {\bf 904-905}, 389c (2013)
  doi:10.1016/j.nuclphysa.2013.02.031
  [arXiv:1210.4588 [hep-ph]].
  

\bibitem{Sombun:2018yqh} 
  S.~Sombun, K.~Tomuang, A.~Limphirat, P.~Hillmann, C.~Herold, J.~Steinheimer, Y.~Yan and M.~Bleicher,
  arXiv:1805.11509 [nucl-th].
  
\bibitem{Gyulassy:1982pe} 
  M.~Gyulassy, K.~Frankel and E.~a.~Remler,
  Nucl.\ Phys.\ A {\bf 402}, 596 (1983).
  doi:10.1016/0375-9474(83)90222-1
  

  
\bibitem{Danielewicz:1991dh} 
  P.~Danielewicz and G.~F.~Bertsch,
  Nucl.\ Phys.\ A {\bf 533}, 712 (1991).
  doi:10.1016/0375-9474(91)90541-D

\bibitem{Oh:2009gx} 
  Y.~Oh, Z.~W.~Lin and C.~M.~Ko,
  Phys.\ Rev.\ C {\bf 80}, 064902 (2009)
  doi:10.1103/PhysRevC.80.064902
  [arXiv:0910.1977 [nucl-th]].
  
\bibitem{Longacre:2013apa} 
  R.~S.~Longacre,
  arXiv:1311.3609 [hep-ph].
  
\bibitem{Pang:2018zzo} 
  L.~G.~Pang, H.~Petersen and X.~N.~Wang,
  Phys.\ Rev.\ C {\bf 97}, no. 6, 064918 (2018)
  doi:10.1103/PhysRevC.97.064918
  [arXiv:1802.04449 [nucl-th]].
  
\bibitem{Bernhard:2016tnd} 
  J.~E.~Bernhard, J.~S.~Moreland, S.~A.~Bass, J.~Liu and U.~Heinz,
  Phys.\ Rev.\ C {\bf 94}, no. 2, 024907 (2016)
  doi:10.1103/PhysRevC.94.024907
  [arXiv:1605.03954 [nucl-th]].
  
\bibitem{Schenke:2012wb} 
  B.~Schenke, P.~Tribedy and R.~Venugopalan,
  Phys.\ Rev.\ Lett.\  {\bf 108}, 252301 (2012)
  doi:10.1103/PhysRevLett.108.252301
  [arXiv:1202.6646 [nucl-th]].
  

\bibitem{Borsanyi:2012ve} 
  S.~Borsanyi, G.~Endrodi, Z.~Fodor, S.~D.~Katz and K.~K.~Szabo,
  JHEP {\bf 1207}, 056 (2012)
  doi:10.1007/JHEP07(2012)056
  [arXiv:1204.6184 [hep-lat]].
  
\bibitem{Cooper:1974mv} 
  F.~Cooper and G.~Frye,
  Phys.\ Rev.\ D {\bf 10}, 186 (1974).
  doi:10.1103/PhysRevD.10.186

\bibitem{Weil:2016zrk} 
  J.~Weil {\it et al.},
  Phys.\ Rev.\ C {\bf 94}, no. 5, 054905 (2016)
  doi:10.1103/PhysRevC.94.054905
  [arXiv:1606.06642 [nucl-th]].
  
\bibitem{Patrignani:2016xqp} 
  C.~Patrignani {\it et al.} [Particle Data Group],
  Chin.\ Phys.\ C {\bf 40}, no. 10, 100001 (2016).
  doi:10.1088/1674-1137/40/10/100001

\bibitem{Steinberg:2018jvv} 
  V.~Steinberg, J.~Staudenmaier, D.~Oliinychenko, F.~Li, Ö.~Erkiner and H.~Elfner,
  arXiv:1809.03828 [nucl-th].
  
\bibitem{Bass:1998ca} 
  S.~A.~Bass {\it et al.},
  Prog.\ Part.\ Nucl.\ Phys.\  {\bf 41}, 255 (1998)
  [Prog.\ Part.\ Nucl.\ Phys.\  {\bf 41}, 225 (1998)]
  doi:10.1016/S0146-6410(98)00058-1
  [nucl-th/9803035].



  
\bibitem{SAID_database}
http://gwdac.phys.gwu.edu

\bibitem{Arndt:1994bs} 
  R.~A.~Arndt, I.~I.~Strakovsky and R.~L.~Workman,
  Phys.\ Rev.\ C {\bf 50}, 1796 (1994)
  doi:10.1103/PhysRevC.50.1796
  [nucl-th/9407032].

\bibitem{Carlson:1973}
  R.~F.~Carlson {\it et at.},
  Lett.\ Nuovo\ Cim. {\bf 8}, 319 (1973)
  
\bibitem{Sibirtsev:2006yw} 
  A.~Sibirtsev, J.~Haidenbauer, S.~Krewald and U.~G.~Meissner,
  J.\ Phys.\ G {\bf 32}, R395 (2006)
  doi:10.1088/0954-3899/32/11/R02
  [nucl-th/0608028].
  
\bibitem{Giacomelli:1972vb} 
  G.~Giacomelli {\it et al.},
  Nucl.\ Phys.\ B {\bf 37}, 577 (1972).
  doi:10.1016/0550-3213(72)90520-2
  
\bibitem{Bugg:1968zz} 
  D.~V.~Bugg {\it et al.},
  Phys.\ Rev.\  {\bf 168}, 1466 (1968).
  doi:10.1103/PhysRev.168.1466
  
\bibitem{Bizzarri:1973sp} 
  R.~Bizzarri, P.~Guidoni, F.~Marcelja, F.~Marzano, E.~Castelli and M.~Sessa,
  Nuovo Cim.\ A {\bf 22}, 225 (1974).
  doi:10.1007/BF02813436
  
\bibitem{Abelev:2012wca} 
  B.~Abelev {\it et al.} [ALICE Collaboration],
  Phys.\ Rev.\ Lett.\  {\bf 109}, 252301 (2012)
  doi:10.1103/PhysRevLett.109.252301
  [arXiv:1208.1974 [hep-ex]].
  
\bibitem{Acharya:2017dmc} 
  S.~Acharya {\it et al.} [ALICE Collaboration],
  Eur.\ Phys.\ J.\ C {\bf 77}, no. 10, 658 (2017)
  doi:10.1140/epjc/s10052-017-5222-x
  [arXiv:1707.07304 [nucl-ex]].
  
\bibitem{Danielewicz:1992pei} 
  P.~Danielewicz and P.~Schuck,
  Phys.\ Lett.\ B {\bf 274}, 268 (1992).
  doi:10.1016/0370-2693(92)91985-I


\bibitem{Scheibl:1998tk} 
  R.~Scheibl and U.~W.~Heinz,
  Phys.\ Rev.\ C {\bf 59}, 1585 (1999)
  doi:10.1103/PhysRevC.59.1585
  [nucl-th/9809092].

\bibitem{Oliinychenko:2018odl} 
  D.~Oliinychenko, L.~G.~Pang, H.~Elfner and V.~Koch,
  arXiv:1812.06225 [hep-ph],
  to be published in MDPI Proceedings

\bibitem{Yu:2017bxv} 
  N.~Yu [STAR Collaboration],
  Nucl.\ Phys.\ A {\bf 967}, 788 (2017)
  doi:10.1016/j.nuclphysa.2017.06.046
  [arXiv:1704.04335 [nucl-ex]].
  

\bibitem{Dingwei_Zhang}
  Dingwei~Zhang [STAR Collaboration]
  Poster at Quark Matter 2018 conference
  
\bibitem{Xu:2004mz} 
  Z.~Xu and C.~Greiner,
  Phys.\ Rev.\ C {\bf 71}, 064901 (2005)
  doi:10.1103/PhysRevC.71.064901
  [hep-ph/0406278].
  
\bibitem{Seifert:2017oyb} 
  E.~Seifert and W.~Cassing,
  Phys.\ Rev.\ C {\bf 97}, no. 2, 024913 (2018)
  doi:10.1103/PhysRevC.97.024913
  [arXiv:1710.00665 [hep-ph]].
  
\bibitem{Pan:2014caa} 
  Y.~Pan and S.~Pratt,
  Phys.\ Rev.\ C {\bf 89}, no. 4, 044911 (2014).
  doi:10.1103/PhysRevC.89.044911
  
  
\bibitem{Bashkanov:2015xsa} 
  M.~Bashkanov, H.~Clement and T.~Skorodko,
  Eur.\ Phys.\ J.\ A {\bf 51}, no. 7, 87 (2015)
  doi:10.1140/epja/i2015-15087-x
  [arXiv:1502.07156 [nucl-ex]].
 
\end{thebibliography}
\end{document}